\newif\ifonecolumn
\onecolumnfalse

\newif\ifarxiv
\arxivtrue

\newif\iftwocolumn
\ifonecolumn
    \twocolumnfalse
\else
    \twocolumntrue
\fi

\ifonecolumn
    \documentclass[12pt, draftclsnofoot, onecolumn]{IEEEtran}
    
    \usepackage[font=footnotesize]{subcaption}
    \usepackage{mwe}
    \usepackage[export]{adjustbox}
    \usepackage{microtype} 
    \usepackage{relsize}

    \newcommand\spaceunderfig{-40pt}

\else
    \documentclass[journal, twocolumn]{IEEEtran}
    \newcommand\spaceunderfig{-10pt}
\fi

\newcommand\speciallinespacing{1.00}

\usepackage{graphicx}
\usepackage{cite}
\usepackage{multirow, array}
\usepackage[cmex10]{amsmath}
\usepackage{amssymb}
\usepackage{cases}
\usepackage[font=footnotesize]{caption}
\usepackage{siunitx}
\usepackage{xcolor}
\usepackage{placeins}
\usepackage{xspace}
\usepackage{textcomp}
\usepackage{import}
\usepackage[update,prepend]{epstopdf}
\usepackage{microtype}
\usepackage[nonumberlist]{glossaries}
\usepackage{bm}
\usepackage[super]{nth}
\usepackage{relsize}
\usepackage{url}

\newcommand{\MKcolor}[1]{\textcolor{black}{#1}}

\newacronym{5G}{5G}{fifth generation}
\newacronym{ACF}{ACF}{autocorrelation function}
\newacronym{AR}{AR}{auto-regressive}
\newacronym{AN}{AN}{access node}
\newacronym{AoA}{AoA}{angle of arrival}
\newacronym{AWGN}{AWGN}{additive white Gaussian noise}
\newacronym{BS}{BS}{base station}
\newacronym{BIC}{BIC}{Bayesian information criterion}
\newacronym{BPSK}{BPSK}{binary phase shift keying}
\newacronym{CL}{CL}{centroid localization}
\newacronym{CR}{CR}{cognitive radio}
\newacronym{CRB}{CRB}{Cramer-Rao bound}
\newacronym{CSI}{CSI}{channel state information}
\newacronym{CSIT}{CSIT}{channel state information at transmitter}
\newacronym{CWNA}{CWNA}{continuous white noise acceleration}
\newacronym{D2D}{D2D}{device-to-device}
\newacronym{DBS}{DBS}{different beamwidth sectors}
\newacronym{DCAA}{DCAA}{digitally controlled antenna array}
\newacronym{DFU}{DFU}{{}\acrshort{DoA} fusion}
\newacronym{DL}{DL}{downlink}
\newacronym[firstplural=directions of arrival (DoAs)]{DoA}{DoA}{direction of arrival}
\newacronym{DS-CDMA}{DC-CDMA}{direct sequence code division multiple access}
\newacronym{E911}{E911}{enhanced 911}
\newacronym{EADF}{EADF}{effective aperture distribution function}
\newacronym{EBS}{EBS}{equal beamwidth sectors}
\newacronym{EKF}{EKF}{extended Kalman filter}
\newacronym{ESA}{ESA}{equal sector antenna}
\newacronym{ESPAR}{ESPAR}{electronically steerable parasitic array radiator}
\newacronym{EOTD}{E-OTD}{enhanced observed time difference}
\newacronym{EW}{EW}{equal weighting}
\newacronym{FCC}{FCC}{federal communications commission}
\newacronym{FFT}{FFT}{fast Fourier transform}
\newacronym[firstplural=Fisher information matrices (FIM)]{FIM}{FIM}{Fisher information matrix}
\newacronym{GMM}{GMM}{Gaussian mixture model}
\newacronym{GNSS}{GNSS}{global navigation satellite system}
\newacronym{GPS}{GPS}{global positioning system}
\newacronym{GSM}{GSM}{global system for mobile communications}
\newacronym{ICI}{ICI}{inter-carrier-interference}
\newacronym{ICMP}{ICMP}{internet control message protocol}
\newacronym{IoT}{IoT}{Internet-of-Things}
\newacronym{ISD}{ISD}{inter-site distance}
\newacronym{ISI}{ISI}{inter-symbol-interference}
\newacronym{ITS}{ITS}{intelligent traffic system}
\newacronym{ITU}{ITU}{international telecommunication union}
\newacronym{KF}{KF}{Kalman filter}
\newacronym{LMMSE}{LMMSE}{linear minimum mean-square error}
\newacronym{LoS}{LoS}{line-of-sight}
\newacronym{LS}{LS}{least squares}
\newacronym{LTE}{LTE}{long term evolution}
\newacronym{LWA}{LWA}{leaky-wave antenna}
\newacronym{MAP}{MAP}{maximum a posteriori} 
\newacronym{METIS}{METIS}{Mobile and wireless communications enablers for the twenty-twenty information society}
\newacronym{MGSCM}{MGSCM}{METIS geometry-based stochastic channel model}
\newacronym{MIMO}{MIMO}{multiple-input multiple-output}
\newacronym{MSE}{MSE}{mean-squared error}
\newacronym{ML}{ML}{maximum likelihood}
\newacronym{MLE}{MLE}{maximum likelihood estimator}
\newacronym{MMSE}{MMSE}{minimum mean-square error}
\newacronym{MU-MIMO}{MU-MIMO}{multiuser multiple-input-multiple-output}
\newacronym{MVUE}{MVUE}{minimum variance unbiased estimator}
\newacronym{NLoS}{NLoS}{non-line-of-sight}
\newacronym{ppm}{ppm}{parts per million}
\newacronym{OFDM}{OFDM}{orthogonal frequency-division multiplexing}
\newacronym{OFDMA}{OFDMA}{orthogonal frequency-division multiple access}
\newacronym{ON}{ON}{observing node}
\newacronym{OTDoA}{OTDoA}{observed time difference of arrival}
\newacronym{PDF}{PDF}{probability density function}
\newacronym{PDoA}{PDoA}{phase difference of arrival}
\newacronym{RToA}{RToA}{round-trip time of arrival}
\newacronym{PPMCC}{PPMCC}{Pearson product-moment correlation coefficient}
\newacronym{PU}{PU}{primary user}
\newacronym{PW}{PW}{power weighting}
\newacronym{RMSE}{RMSE}{root-mean-squared error}
\newacronym{RRMSE}{RRMSE}{relative root-mean-squared error}
\newacronym{RSS}{RSS}{received signal strength}
\newacronym{RRM}{RRM}{radio resource management}
\newacronym{RF}{RF}{radio frequency}
\newacronym{RFI}{RFI}{radio frequency interference}
\newacronym{SBS}{SBS}{switched-beam system}
\newacronym{SDE}{SDE}{sector-pair \acrshort{DoA} estimation}
\newacronym{SIMO}{SIMO}{single-input-multiple-output}
\newacronym{SINR}{SINR}{signal-to-interference-plus-noise ratio}
\newacronym{SLS}{SLS}{simplified least squares}
\newacronym{SNR}{SNR}{signal-to-noise ratio}
\newacronym{SSP}{SSP}{side-sector suppression}
\newacronym{SSL}{SSL}{sector selection}
\newacronym{STD}{STD}{standard deviation}
\newacronym{SU}{SU}{secondary user}
\newacronym{TCP}{TCP}{transmission control protocol}
\newacronym{TDoA}{TDoA}{time difference of arrival}
\newacronym{TN}{TN}{target node}
\newacronym{ToA}{ToA}{time of arrival}
\newacronym{ToF}{ToF}{time of flight}
\newacronym{TSLS}{TSLS}{three-stage SLS}
\newacronym{TTI}{TTI}{transmit time interval}
\newacronym{UDN}{UDN}{ultra-dense network}
\newacronym{UL}{UL}{uplink}
\newacronym{UTDoA}{U-TDoA}{uplink-time difference of arrival}
\newacronym{UMi}{UMi}{urban micro}
\newacronym{UN}{UN}{user node}
\newacronym{VW}{VW}{variance weighting}
\newacronym{WCL}{WCL}{weighted centroid localization}
\newacronym{WLAN}{WLAN}{wireless local area network}
\newacronym{wrt}{wrt}{with respect to}
\newacronym{WSS}{WSS}{wide sense stationary}
\newif\ifpublicversion
\publicversionfalse

\newif\ifsmallfigures
\smallfiguresfalse

\renewcommand{\vec}[1]{\ensuremath{\boldsymbol{\mathbf{#1}}}}
\newcommand{\mat}[1]{\ensuremath{\boldsymbol{\mathbf{#1}}}}

\newcommand*{\eye}[1][]{%
\ifthenelse{\equal{#1}{}}{\ensuremath{\mat{I}}}{\ensuremath{\mat{I}_{#1}}}%
}
\newcommand*{\zeros}[1][]{%
\ifthenelse{\equal{#1}{}}{\ensuremath{\mat{0}}}{\ensuremath{\mat{0}_{#1}}}%
}
\newcommand*{\ones}[1][]{%
\ifthenelse{\equal{#1}{}}{\ensuremath{\mat{1}}}{\ensuremath{\mat{1}_{#1}}}%
}

\newcommand*{\transpose}{\ensuremath{^ \textrm{T}}}
\newcommand*{\conT}{\ensuremath{^{\textrm{*}}}}

\newcommand*{\est}[1]{\ensuremath{\hat{#1}}}

\usepackage{pifont}
\newcolumntype{C}[1]{>{\centering\let\newline\\\arraybackslash\hspace{0pt}}m{#1}}
\newcolumntype{L}[1]{>{\raggedright\let\newline\\\arraybackslash\hspace{0pt}}m{#1}}

\newcommand{\doa}{\ensuremath{\gls{doa}}}

\newcommand{\sdoa}[1][]{%
\ifthenelse{\equal{#1}{}}{\ensuremath{\ensuremath{\sigma_{\doa}}}\xspace}{\ensuremath{\sigma_{\doa, #1}}\xspace}%
}

\ifpublicversion
	
	\newcommand*{\REFC}[1][REF]{\xspace}
\else
	
	\newcommand*{\REFC}[1][REF]{~{\color{blue}[#1]}\xspace}
\fi

\newcommand*{\rfig}[1]{Fig.~\ref{#1}}

\ifpublicversion
	\newcommand*{\SCOM}[1]{}

	\newcommand*{\ACOM}[1]{}
\else
	\newcommand*{\SCOM}[1]{{\color{blue} \it #1}\xspace}

	\newcommand*{\ACOM}[1]{{\color{red} \it #1}\xspace}
\fi

\newcommand{\scalethanks}[1]{

{\linespread{\speciallinespacing}
\let\thefootnote\relax\footnotetext{
\footnotesize
}}}

\begin{document}

\bstctlcite{IEEEexample:BSTcontrol}


\title{Joint Device Positioning and Clock Synchronization in 5G Ultra-Dense Networks}
\author{Mike~Koivisto, M\'ario~Costa, Janis~Werner, Kari~Heiska, Jukka~Talvitie, Kari~Lepp\"anen, Visa~Koivunen and Mikko~Valkama}%


\ifarxiv
\else
\markboth{IEEE Transactions on Wireless Communications}%
{ }
\fi

\maketitle

\vspace{-40pt}

{\linespread{\speciallinespacing}
\let\thefootnote\relax\footnotetext{
\footnotesize
M.~Koivisto, J.~Werner, J.~Talvitie and M.~Valkama are with the Department of Electronics and Communications Engineering, Tampere University of Technology, FI-33101 Tampere, Finland (email: mike.koivisto@tut.fi).
    
M.~Costa, K.~Heiska, and K.~Lepp\"anen are with Huawei Technologies Oy (Finland) Co., Ltd, Helsinki 00180, Finland.

V.~Koivunen is with the Department of Signal Processing and Acoustics, Aalto University, FI-02150 Espoo, Finland.

This work was supported by the Finnish Funding Agency for Technology and Innovation (Tekes), under the projects \textquotedblleft{5G Networks and Device Positioning\textquotedblright}, and \textquotedblleft{Future Small-Cell Networks using Reconfigurable Antennas\textquotedblright}.

Preliminary work addressing a limited subset of initial results was presented at IEEE Global Communications Conference (GLOBECOM), San Diego, CA, USA, December 2015~\cite{werner_joint_2015}.

Online video material available at \texttt{\url{http://www.tut.fi/5G/TWC16/}}.

\ifarxiv
This work has been submitted to the IEEE for possible publication. This is the revised version of the original work and it is currently under review. Copyright may be transferred without notice, after which this version may no longer be accessible.
\fi
}}

\begin{abstract}
\MKcolor{In this article, we address the prospects and key enabling technologies for highly efficient and accurate device positioning and tracking in \gls{5G} radio access networks.} Building on the premises of \glspl{UDN} as well as on the adoption of multicarrier waveforms and antenna arrays in the \glspl{AN}, we first formulate \gls{EKF}-based solutions for \MKcolor{computationally efficient} joint estimation and tracking of the \gls{ToA} and \gls{DoA} of the \glspl{UN} using \gls{UL} reference signals. Then, a second EKF stage is proposed in order to fuse the individual \gls{DoA}/\gls{ToA} estimates from \MKcolor{one or several} \glspl{AN} into a \gls{UN} position estimate. 
\MKcolor{Since all the processing takes place at the network side, the computing complexity and energy consumption at the UN side are kept to a minimum.}
The cascaded \glspl{EKF} proposed in this article also take into account the unavoidable relative clock offsets between \glspl{UN} and \glspl{AN}, such that reliable clock synchronization of the access-link is obtained as a valuable by-product. The proposed cascaded \gls{EKF} scheme is then revised and extended to more general and challenging scenarios where not only the \glspl{UN} have clock offsets against the network time, but also the \glspl{AN} themselves are not mutually synchronized in time. Finally, comprehensive performance evaluations of the proposed solutions on a realistic \gls{5G} network setup, building on the METIS project based outdoor Madrid map model together with complete ray tracing based propagation modeling, are provided. The obtained results clearly demonstrate that by using the developed methods, sub-meter scale positioning and tracking accuracy of moving devices is indeed technically feasible in future 5G radio access networks operating at sub-\SI{6}{GHz} frequencies, despite the realistic assumptions related to clock offsets and potentially even under unsynchronized network elements.
\end{abstract}
\vspace{-5pt}

%
\IEEEpeerreviewmaketitle

\section{Introduction}

5G mobile communication networks are expected to provide major enhancements in terms of, e.g., peak data rates, area capacity, \gls{IoT} support and end-to-end latency, compared to the existing radio systems~\cite{osseiran_scenarios_2014,5g_forum_5g_2015,ngmn_alliance_5g_2014}. In addition to such improved communication features, 5G networks are also expected to enable highly-accurate device or \gls{UN} positioning,  if designed properly~\cite{5g_forum_5g_2015,ngmn_alliance_5g_2014}. Compared to the existing radio positioning approaches, namely \gls{EOTD}~\cite{roxin_survey_2007,sun_signal_2005}, \gls{UTDoA}~\cite{roxin_survey_2007}, \gls{OTDoA}~\cite{medbo_propagation_2009}, which all yield positioning accuracy in the range of few tens of meters, as well as to \gls{GPS}~\cite{dardari_indoor_2015} or WiFi fingerprinting~\cite{liu_push_2012} based solutions in which the accuracy is typically in the order of 3-5 meters at best, the positioning accuracy of 5G networks is expected to be in the order of one meter or even below~\cite{5g_forum_5g_2015,ngmn_alliance_5g_2014,5g-ppp_2015}. Furthermore, as shown in our preliminary work in~\cite{werner_joint_2015}, the positioning algorithms can be carried out at the network side, thus implying a highly energy-efficient approach from the devices’ perspective. 

Being able to estimate and track as well as predict the device positions in the radio network is generally highly beneficial from various perspectives. For one, this can enable location-aware communications~\cite{di_taranto2014,hakkarainen_high-efficiency_2015} and thus contribute to improve the actual core 5G network communications functionalities as well as the radio network operation and management. Concrete examples where device position information can be utilized include network-enabled \gls{D2D} communications~\cite{kuruvatti_robustness_2015}, positioning of a large number of IoT sensors, content prefetching, proactive \gls{RRM} and mobility management~\cite{hakkarainen_high-efficiency_2015}. \MKcolor{Furthermore, cm-wave based 5G radio networks could assist and relax the device discovery problem~\cite{shokri-ghadikolaei_beam-searching_2015} in mm-wave radio access systems. In particular, the cm-wave based system could provide the UN position information that is needed for designing the transmit and receive beamformers for the mm-wave access~\cite{kela_location_2016}.} Continuous and highly accurate network-based positioning, either in 2D or even 3D, is also a central enabling technology for self-driving cars, \glspl{ITS} and collision avoidance, drones as well as other kinds of autonomous vehicles and robots which are envisioned to be part of not only the future factories and other production facilities but the overall future society within the next 5-10 years~\cite{fettweis_tactile_2014}.

In this article, building on the premises of ultra-dense 5G networks~\cite{osseiran_scenarios_2014,5g_forum_5g_2015,ngmn_alliance_5g_2014,bhushan_network_2014}, \MKcolor{we develop enabling technical solutions that facilitate obtaining and providing device location information in 5G systems with both high-accuracy and low power consumption at the user devices.} We focus on the connected vehicles type of scenario, which is identified, e.g., in~\cite{ngmn_alliance_5g_2014,5g-ppp_2015} as one key application and target for future 5G mobile communications, with a minimum of 2000 connected vehicles per square kilometer and at least \SI{50}{Mbps} per-car \gls{DL} rate~\cite{ngmn_alliance_5g_2014}. In general, \glspl{UDN} are particularly well suited for network-based \gls{UN} positioning \MKcolor{as illustrated in Fig.~\ref{fig:system_model}}. As a result of the high density of \glspl{AN}, \glspl{UN} in such networks are likely to have a \gls{LoS} towards multiple \glspl{AN} for most of the time even in demanding propagation environments. Such \gls{LoS} conditions alone are already a very desirable property in positioning systems~\cite{wymeersch2009}. Furthermore, the 5G radio networks are also expected to operate with very short radio frames, the corresponding sub-frames or \glspl{TTI} being in the order of \SIrange{0.1}{0.5}{ms}, as described, e.g., in \cite{Kela15,Lahetkangas13}. These short sub-frames generally include \gls{UL} pilots that are intended for \gls{UL} channel estimation and also utilized for \gls{DL} precoder design. In addition, these \gls{UL} pilots can be then also exploited for network-centric \gls{UN} positioning and tracking. More specifically, \glspl{AN} that are in \gls{LoS} with a UN can use the UL pilots to estimate the \gls{ToA} efficiently. Due to the very broad bandwidth waveforms envisioned in 5G, in the order of \SI{100}{MHz} and beyond~\cite{Kela15,Levanen14}, the \glspl{ToA} can generally be estimated with a very high accuracy. Since it is moreover expected that \glspl{AN} are equipped with antenna arrays, \gls{LoS}-\glspl{AN} can also estimate the \gls{DoA} of the incoming \gls{UL} pilots. Then, through the fusion of \gls{DoA} and \gls{ToA} estimates across one or more \glspl{AN}, highly accurate \gls{UN} position estimates can be obtained, and tracked over time, as it will be demonstrated in this article.

More specifically, the novelty and technical contributions of this article are the following. Building on~\cite{SRK09} and our preliminary work in~\cite{werner_joint_2015}, we first formulate a computationally efficient \gls{EKF} for joint estimation and tracking of \gls{DoA} and \gls{ToA}. Such an \gls{EKF} is the core processing engine at individual \glspl{AN}. Then, for efficient fusion of the \gls{DoA}/\gls{ToA} estimates of multiple \glspl{AN} into a device position estimate, a second \gls{EKF} stage is proposed as depicted in Fig.~\ref{fig:cascaded_ekf}. \MKcolor{Compared to existing literature, such as~\cite{aidala_kalman_1979,navarro_frequency_2011,navarro_toa_2007, yousefi_mobile_2015}, the cascaded \glspl{EKF} proposed in this paper also take into account the unavoidable relative clock offsets among \glspl{UN} and receiving \glspl{AN}. Hence, accurate clock offset estimates are obtained as a by-product.} This makes the proposed approach much more realistic, compared to earlier reported work, while being able to estimate the \gls{UN} clock offsets has also a high value of its own. Then, as another important contribution, we also develop a highly accurate cascaded \gls{EKF} solution for scenarios where not only the \glspl{UN} have clock offsets against the network time, but also the \glspl{AN} themselves are not mutually synchronized in time. Such an \gls{EKF}-based fusion solution provides an advanced processing engine inside the network where the \gls{UN} positions and clock offsets as well as valuable \gls{AN} clock offsets are all estimated and tracked. \MKcolor{As a concrete example, in \gls{OTDoA}-based positioning in LTE, the typical value for the clock offsets among the \glspl{AN} is assumed to be less than \SI{0.1}{\micro s}~\cite[Table~8-1]{Fischer14}. Furthermore, the expected timing misalignment requirement for future 5G small-cell networks is less than \SI{0.5}{\micro s}~\cite{mogensen_5g_2013}, thus giving us a concrete quantitative reference regarding network synchronization.}

\iftwocolumn
{\linespread{\speciallinespacing}
\begin{figure}
    \centering
    \includegraphics[width=3.12in]{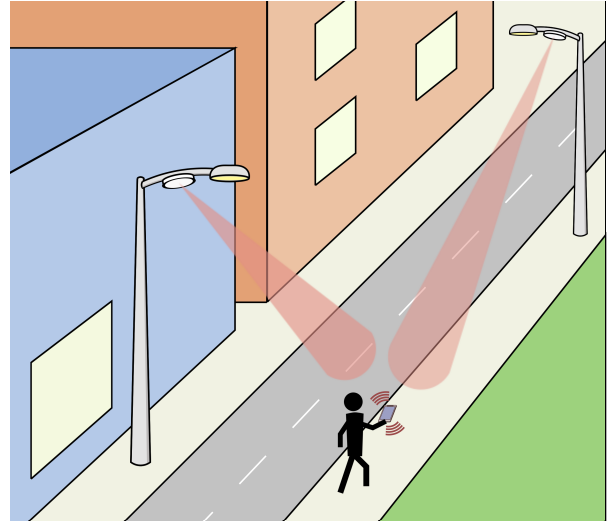}
    \caption{Positioning in $5$G \glspl{UDN}. Multiantenna \glspl{AN} and multicarrier waveforms make it possible to estimate and track the position of the \gls{UN} with high-accuracy by relying on \gls{UL} reference signals, used primarily for \gls{DL} precoder calculation.}
    \label{fig:system_model}
    \vspace{\spaceunderfig}
\end{figure}
}
\fi

To the best of authors’ knowledge, such solutions have not been reported earlier in the existing literature. For generality, we note that a \gls{MLE} for joint \gls{UN} localization and network synchronization has been proposed in~\cite{jean_passive_2014}. However, such an algorithm is a batch solution and does not provide sequential estimation of the UN position and synchronization parameters needed in mobile scenarios and dynamic propagation environments. In practice, both the UN position and synchronization parameters are time-varying. \MKcolor{Moreover, the work in~\cite{jean_passive_2014,yousefi_sensor_2015} focus on \gls{ToA} measurements only, thus requiring fusing the measurements from a larger amount of \glspl{AN} than that needed in our approach.} Hence, this article may be understood as a considerable extension of the work in~\cite{jean_passive_2014} where both \gls{ToA} and \gls{DoA} measurements are taken into account for sequential estimation and tracking of \gls{UN} position and network synchronization. A final contribution of this article consists of providing a vast and comprehensive performance evaluation of the proposed solutions in a realistic 5G network setup, building on the METIS project Madrid map model~\cite{metis_simulation_2013}. The network is assumed to be operating at \SI{3.5}{GHz} band, and the \gls{MIMO} channel propagation for the \gls{UL} pilot transmissions is modeled by means of a ray tracing tool where all essential propagation paths are emulated. In the performance evaluations, various parameters such as the \gls{AN} \gls{ISD} and \gls{UL} pilot spacing in frequency are varied. \MKcolor{In addition, the positioning and synchronization performance is evaluated by fusing the estimated DoA and ToA measurements from a varying and realistic number of LoS-ANs. It should be noted that numerical results considering imperfect LoS-detection are also provided in this paper}. The obtained results demonstrate that sub-meter scale positioning accuracy is indeed technically feasible in future 5G radio access networks, even under the realistic assumptions related to time-varying clock offsets. The results also indicate that the proposed \gls{EKF}-based solutions provide highly-accurate clock offset estimates not only for the \glspl{UN} but also across network elements, which contains high value on its own, namely for synchronization of 5G \glspl{UDN}.

\iftwocolumn
{\linespread{\speciallinespacing}
\begin{figure}
    \centering
    \includegraphics[width=3.12in]{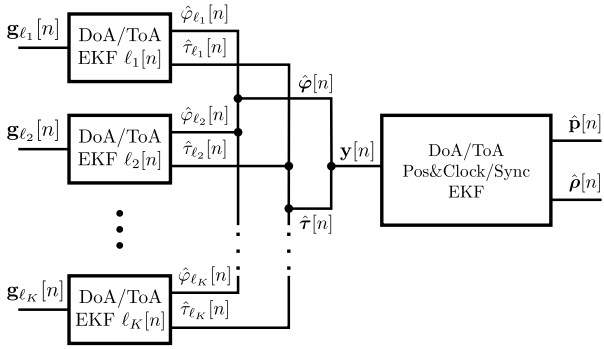}
    \caption{Cascaded \acrfullpl{EKF} for joint \acrfull{UN} positioning and network clock synchronization. The \gls{DoA}/\gls{ToA} \glspl{EKF} operate in a distributed manner at each \gls{AN} while the Pos\&Clock/Sync \glspl{EKF} operate in a central-unit fusing the azimuth DoA and ToA measurements of $K[n]$ ANs.}
    \label{fig:cascaded_ekf}
    \vspace{\spaceunderfig}
\end{figure}
}
\fi

The rest of the article is organized as follows. In Section~\ref{sec:system_model}, we describe the basic system model, including the assumptions related to the ultra-dense 5G network, antenna array models in the \glspl{AN} and the clock offset models adopted for the \gls{UN} devices and network elements. 
The proposed solutions for joint \gls{DoA}/\gls{ToA} estimation and tracking at individual \glspl{AN} as well as for joint UN position and clock offset estimation and tracking in the network across \glspl{AN} are all described in Section \ref{sec:DoA-ToA_EKF}. In Section \ref{sec:syncpos}, we provide the extension to the case of unsynchronized network elements, and describe the associated \gls{EKF} solutions for estimation and tracking of all essential parameters including the mutual clock offsets of \glspl{AN}. Furthermore, the propagation of universal network time is shortly addressed. In Section \ref{sec:results}, we report the results of extensive numerical evaluations in realistic 5G network context, while also comparing the results to those obtained using earlier prior art. Finally, conclusions are drawn in Section \ref{sec:conclusion}.

\section{System Model} \label{sec:system_model}


%
%

\subsection{5G Ultra-Dense Networks and Positioning Engine}
We consider an \gls{UDN} where the \glspl{AN} are equipped with multiantenna transceivers. The \glspl{AN} are deployed below rooftops and have a maximum \gls{ISD} of around \SI{50}{m}; see Fig.~\ref{fig:system_model}. The \gls{UN} transmits periodically \gls{UL} reference signals in order to allow for multiuser MIMO (MU-MIMO) schemes based on \gls{CSIT}. The \gls{UL} reference signals are assumed to employ a multicarrier waveform such as \gls{OFDM}, in the form of \gls{OFDMA} in a multiuser network. These features are widely accepted to be part of $5$G \gls{UDN} developments, as discussed, e.g., in \cite{osseiran_scenarios_2014,5g_forum_5g_2015,ngmn_alliance_5g_2014,5g-ppp_2015,metis_simulation_2013}, and in this paper we take advantage of such a system in order to provide and enable high-efficiency UN positioning and network synchronization.

In particular, the multiantenna capabilities of the \glspl{AN} make it possible to estimate the \gls{DoA} of the \gls{UL} reference signals while employing multicarrier waveforms allows one to estimate the \gls{ToA} of such \gls{UL} pilots. The position of the \gls{UN} is then obtained with the proposed \gls{EKF} by fusing the \gls{DoA} and \gls{ToA} estimates from multiple \glspl{AN}, given that such \glspl{AN} are in \gls{LoS} condition with the \gls{UN}. In fact, the \gls{LoS} probability in \glspl{UDN} comprised of \glspl{AN} with a maximum \gls{ISD} of \SI{50}{m} is very high, e.g., $0.8$ in the stochastic channel model descibed in \cite{metis_channel_2015,3GPP_channel} and already around $0.95$ for an \gls{ISD} of \SI{40}{m}. Note that the \gls{LoS}/\gls{NLoS} condition of a \gls{UN}-\gls{AN} link may be determined based on the Rice factor of the received signal strength, as described, e.g., in \cite{BGTV07}. \MKcolor{For the sake of generality, we analyse the performance of the proposed methods under both perfect and imperfect LoS-detection scenarios.}

In this paper, we focus on 2D positioning ($xy$-plane only) and assume that the locations of the \glspl{AN} are fully known. However, the extension of the \glspl{EKF} proposed here to 3D positioning is straightforward. \MKcolor{We also note that the \gls{EKF}-based methods proposed in this paper can be used for estimating the positions of the \glspl{AN} as well, given that only a few \glspl{AN} are surveyed. In practice, such an approach could decrease the deployment cost and time of a \gls{UDN}.} We further assume two different scenarios for synchronization within a network. First, \glspl{UN} are assumed to have unsynchronized {\linespread{\speciallinespacing}\footnote{We assume that the timing and frequency synchronization needed for avoiding \gls{ICI} and \gls{ISI} has been achieved. Such an assumption is similar to that needed in \gls{OFDM} based wireless systems in order to decode the received data symbols. \MKcolor{In principle, \gls{ICI} can be understood as a systematic error in the measurement model, and may thus be taken into account in the proposed \glspl{EKF} simply by increasing the measurement noise covariance~\cite[Ch.3]{grooves13}. However, throughout this paper, ICI is assumed to be negligible while more rigorous methods to account for ICI are left for future work. }}} clocks whereas the clocks within \glspl{AN} are assumed to be synchronized among each other. Second, not only the clock of a \gls{UN} but also the clocks within \glspl{AN} are assumed to be unsynchronized. For the sake of simplicity, we make an assumption that the clocks within \glspl{AN} are phase-locked in the second scenario, i.e., the clock offsets of the \glspl{AN} are essentially not varying with respect to the actual time. Completely synchronized as well as phase-locked clocks can be adjusted using a reference time from, e.g., \gls{GPS}, or by communicating a reference signal from a central-entity of the network to the \glspl{AN}, but these methods surely increase the signaling overhead.

\subsection{Channel Model for DoA/ToA Estimation and Tracking} \label{sec:ChannelModel_tracking}

The channel model employed by the proposed \gls{EKF} for estimating and tracking the \gls{DoA}/\gls{ToA} parameters comprises a single dominant path. It is important to note that a detailed ray tracing based channel model is then used in all the numerical results for emulating the estimated channel frequency responses at the \glspl{AN}. However, the \gls{EKF} proposed in this paper fits a single-path model to the estimated multipath channel. The motivation for such an approach is twofold. Firstly, the typical Rice factor in \glspl{UDN} is \SIrange{10}{20}{dB}~\cite{metis_channel_2015,3GPP_channel}. Secondly, the resulting \gls{EKF} is computationally more efficient than the approach of estimating and tracking multiple propagation paths~\cite{SRK09}. This method thus allows for reduced computing complexity, while still enabling high-accuracy positioning and tracking, as will be shown in the evaluations.

In particular, the \glspl{EKF} proposed in Section \ref{sec:doa_toa_ekf} exploit the following model for the \gls{UL} \gls{SIMO} multicarrier-multiantenna channel response estimate at an \gls{AN}, obtained using UL reference signals, of the form \cite{Ric05}
\begin{equation} \label{eq:EKF_fitted_model}
    \mathbf{g} \approx \mathbf{B}(\mathbf{\vartheta},\mathbf{\varphi},\mathbf{\tau}) \bm{\gamma} + \mathbf{n},
\end{equation}
where $\mathbf{B}(\mathbf{\vartheta},\mathbf{\varphi},\mathbf{\tau}) \in \mathbb{C}^{\mathcal{M} \times 2}$ and $\bm{\gamma} \in \mathbb{C}^{2\times1}$ denote the polarimetric response of the multicarrier-multiantenna \gls{AN} and the path weights, respectively. Moreover, $\mathbf{n}\in\mathbb{C}^{\mathcal{M} \times 1}$ denotes complex-circular zero-mean white-Gaussian distributed noise with variance $\sigma_n^2$. The dimension of the multichannel vector $\mathbf{g}$ is given by $\mathcal{M} = \mathcal{M}_f \mathcal{M}_{\mathrm{AN}}$, where $\mathcal{M}_f$ and $\mathcal{M}_{\mathrm{AN}}$ denote the number of subcarriers and antenna elements, respectively.

In this paper, either planar or conformal antenna arrays can be employed, and their elements may also be placed non-uniformly. In particular, the polarimetric array response is given in terms of the \gls{EADF} \cite{Ric05,SRK09,CRK2012} as
\ifonecolumn
\else
\begin{align} \label{eq:multichannel_model}
    \mathbf{B}(\vartheta,\varphi,\tau) &= [\mathbf{G}_H \mathbf{d}(\varphi,\vartheta) \otimes \mathbf{G}_f \mathbf{d}(\tau),\nonumber\\
    & \hspace{2cm} \mathbf{G}_V \mathbf{d}(\varphi,\vartheta) \otimes \mathbf{G}_f \mathbf{d}(\tau)],
\end{align}
\fi
where $\otimes$ denotes the Kronecker product. Here, $\mathbf{G}_f\in\mathbb{C}^{\mathcal{M}_f \times \mathcal{M}_f}$ denotes the frequency response of the receivers, and $\mathbf{G}_H\in\mathbb{C}^{\mathcal{M}_{\mathrm{AN}} \times \mathcal{M}_a\mathcal{M}_e}$ and $\mathbf{G}_V\in\mathbb{C}^{\mathcal{M}_{\mathrm{AN}} \times \mathcal{M}_a\mathcal{M}_e}$ denote the \gls{EADF} of the multiantenna AN for a horizontal and vertical excitation, respectively. Also, $\mathcal{M}_a$ and $\mathcal{M}_e$ denote the number of modes, i.e., spatial harmonics, of the array response; see \cite[Ch.2]{Ric05}, \cite{CRK2012} for details. Moreover, $\mathbf{d}(\tau) \in \mathbb{C}^{\mathcal{M}_f \times 1}$ denotes a Vandermonde structured vector given by
\ifonecolumn
\else
\begin{equation} \label{eq:vandermonde_vector}
    \mathbf{d}(\tau) = \left[e^{-j \pi(\mathcal{M}_f-1)f_0 \tau}, \ldots, e^{j \pi(\mathcal{M}_f-1)f_0 \tau}\right]\transpose,
\end{equation}
\fi
where $f_0$ denotes the subcarrier spacing of the adopted multicarrier waveform. Finally, vector $\mathbf{d}(\varphi,\vartheta) \in \mathbb{C}^{\mathcal{M}_a\mathcal{M}_e \times 1}$ is given by \begin{equation}
    \mathbf{d}(\varphi,\vartheta) = \mathbf{d}(\vartheta)\otimes\mathbf{d}(\varphi),
\end{equation}
where $\mathbf{d}(\varphi) \in \mathbb{C}^{\mathcal{M}_a \times 1}$ and $\mathbf{d}(\vartheta)\in \mathbb{C}^{\mathcal{M}_e \times 1}$ have a structure identical to that in \eqref{eq:vandermonde_vector} by using $\pi f_0\tau \rightarrow \vartheta/2$, and similarly for $\varphi$. Note that we have assumed identical \gls{RF}-chains at the multiantenna \gls{AN} and a frequency-flat angular response. Such assumptions are taken for the sake of clarity, and an extension of the \gls{EKF} proposed in Section \ref{sec:doa_toa_ekf} to non-identical RF-chains as well as frequency-dependent angular responses is straightforward but computationally more demanding. Note also that the model in \eqref{eq:multichannel_model} accommodates wideband signals and it is identical to that typically used in space-time array processing \cite{WSK84,BFL13}. Moreover, the array calibration data, represented by the EADF, is assumed to be known or previously acquired by means of dedicated measurements in an anechoic chamber \cite{Ric05,CRK2012}.

\MKcolor{In the \gls{DoA}/\gls{ToA} EKFs, we consider both co-elevation $\vartheta\in[0,\,\pi]$ and azimuth $\varphi\in[0,\,2\pi)$ \gls{DoA} angles even though we eventually fuse only the azimuth \glspl{DoA} $\varphi$ in the 2D positioning and clock offset estimation phase.}
This is due to the challenge of decoupling the azimuth angle from the elevation angle in the \gls{EKF} proposed in Section \ref{sec:doa_toa_ekf} without making further assumptions on the employed array geometry or on the height of the UN. It should be also noted that in an \gls{OFDM}-based system, the parameter $\tau$ as given in \eqref{eq:EKF_fitted_model} (i.e., after the \gls{FFT} operation) denotes the difference between the actual \gls{ToA} (wrt. the clock of the \gls{AN}) of the \gls{LoS} path and the start of the \gls{FFT} window \cite[Ch.3]{PMK07}, \cite{AHK13}. The \gls{ToA} wrt. the clock of the \gls{AN} is then found simply by adding the start-time of the \gls{FFT} window to $\tau$. However, throughout this paper and for the sake of clarify, we will call $\tau$ simply the \gls{ToA}.

\subsection{Clock Models}

In the literature, it is generally agreed that the clock offset $\rho$ is a time-varying quantity due to imperfections of the clock oscillator in the device, see e.g., \cite{wu_clock_2011,kim_tracking_2012,AHK13}. For a measurement period $\Delta t$, the clock offset is typically expressed in a recursive form as \cite{kim_tracking_2012}
\begin{align}
	\rho[n] = \rho[n-1] + \alpha[n] \Delta t \label{eq:clock_offset}
\end{align}
where $\alpha[n]$ is known as the clock skew. Some authors, e.g., the authors in \cite{wu_clock_2011} assume the clock skew to be constant, while some recent research based on measurements suggests that the clock skew can also, in fact, be time-dependent, at least over the large observation period (\SI{1.5}{months}) considered in \cite{kim_tracking_2012}. However, taking the research and measurement results in \cite{kohno_remote_2005,cristea_fingerprinting_2013} into account, where devices are identified remotely based on an estimate of the average clock skew, one could assume that the {\it average clock skew} is indeed constant. This also matches with the measurement results in \cite{kim_tracking_2012}, where the clock skew seems to be fluctuating around a mean value. Nevertheless, the measurements in \cite{kohno_remote_2005,kim_tracking_2012,cristea_fingerprinting_2013} were obtained indoors, i.e., in a temperature controlled environment. However, in practice, environmental effects such as large changes in the ambient temperature affect the clock parameters in the long term \cite{wu_clock_2011}. Therefore, we adopt the more general model \cite{kim_tracking_2012} of a time-varying clock skew, which also encompasses the constant clock skew model as a special case.

The clock skew in \cite{kim_tracking_2012} is modeled as an \gls{AR} process of order $P$. While the measurement results in \cite{kim_tracking_2012} reveal that modeling the clock offset as an \gls{AR} process results in large performance gains compared to a constant clock skew model, an increase of the order beyond $P=1$ does not seem to increase the accuracy of clock offset tracking significantly. \MKcolor{In this paper, the clock skews of the assumed clock oscillators are modeled as an \gls{AR} model of first order according to}
\begin{align}
	\alpha[n] = \beta \alpha[n-1] + \eta[n]\label{eq:clock_skew}
\end{align}
where $|\beta| \leq 1$ is a constant parameter and $\eta[n] \sim \mathcal{N}(0, \sigma_\eta^2)$ is \gls{AWGN}. Note that the joint \gls{DoA}/\gls{ToA} Pos\&Clock \gls{EKF} as well as the joint DoA/ToA Pos\&Sync EKF proposed in Sections \ref{ssec:joint_doa_toa_ekf} and \ref{sec:syncpos}, respectively, may be extended to \gls{AR} processes of higher orders \MKcolor{using, e.g., the state augmentation approach~\cite[Ch.~7.2]{simon_optimal_2006}. In particular, such an approach would be useful for low-grade clock oscillators since their frequency stability is typically poorer than that of medium/high-grade oscillators.}

\section{Cascaded EKFs for Joint \gls{UN} Positioning and \gls{UN} Clock Offset Estimation} \label{sec:DoA-ToA_EKF}
The \gls{EKF} is a widely used estimation method for the \gls{UN} positioning when measurements such as the \glspl{DoA} and \glspl{ToA} are related to the state through a non-linear model, e.g., \cite{navarro_toa_2007}. However, the \gls{UN} positioning is quite often done within the \gls{UN} device which leads to increased energy consumption of the device compared to a network-centric positioning approach \cite{werner_joint_2015}. In this paper, the \gls{UN} positioning together with the \gls{UN} clock offset estimation are done in a network-centric manner using a cascaded \gls{EKF}. \MKcolor{The first part of the cascaded \gls{EKF} consists of tracking the \glspl{DoA} and \glspl{ToA} of a given UN within each \gls{LoS}-\glspl{AN} in a computationally efficient manner}, whereas the second part consists of the joint \gls{UN} positioning and \gls{UN} clock offset estimation, where the DoA and ToA measurements obtained from the first part of the cascaded \gls{EKF} are used. \MKcolor{Since our focus is on 2D positioning, we fuse only the estimated azimuth DoAs and ToAs in the second phase of the cascaded solution.} The structure of the proposed cascaded \gls{EKF} is illustrated in Fig.~ \ref{fig:cascaded_ekf}. \MKcolor{Throughout this paper, we use the same notation as in~\cite{simon_optimal_2006}. Thus, the \textit{a priori} mean and covariance estimates at time instant $n$ are denoted as $\est{\mathbf{s}}^-[n]$ and $\est{\mathbf{P}}^-[n]$, respectively. Similarly, the \textit{a posteriori} mean and covariance estimates, which are obtained after the measurement update phase of the EKF, are denoted as $\est{\mathbf{s}}^+[n]$ and $\est{\mathbf{P}}^+[n]$, respectively.}
%

%
%
\subsection{DoA/ToA Tracking EKF at AN}\label{sec:doa_toa_ekf}

\MKcolor{In this section, an \gls{EKF} for tracking the \gls{DoA} and \gls{ToA} of the \gls{LoS}-path at an AN is formulated, stemming from the work in \cite{SRK09}. However, the formulation of the \gls{EKF} proposed in this paper is computationally more attractive than that in \cite{SRK09} for two reasons. First, the goal in \cite{SRK09} is to have an accurate characterization of the radio channel, and thus all of the significant specular paths need to be estimated and tracked. However, in our work only a single propagation path corresponding to the largest power is tracked.} In addition to computational advantages, the main motivation for using such a model in the \gls{EKF} follows from the fact that the propagation path with largest power typically corresponds to the \gls{LoS} path. This is even more noticeable in \glspl{UDN} where the \gls{AN}-\gls{UN} distance is typically less than \SI{50}{m}, and the Rice-K factor is around \SIrange{10}{20}{dB}~\cite{metis_channel_2015}.
\MKcolor{Second, the \gls{EKF} in \cite{SRK09} tracks a logarithmic parameterization of the path weights (magnitude and phase components) thereby increasing the dimension of the state vector, and consequently the complexity of each iteration of the \gls{EKF}. For \gls{UN} positioning, the path weights can be considered nuisance parameters since the \gls{DoA} and \gls{ToA} suffice in finding the position of the \gls{UN}. It is thus desirable to formulate the \gls{EKF} such that the path weights are not part of the state vector. Hence, the \gls{EKF} proposed in this paper tracks the \gls{DoA} and \gls{ToA} only, thus further decreasing the computational complexity compared to~\cite{SRK09}.} This is achieved by noting that the path weights are linear parameters of the model for the \gls{UL} multicarrier multiantenna channel \cite{Ric05}, and by employing the concentrated log-likelihood function in the derivation of the information-form of the \gls{EKF}~\cite[Ch.6]{simon_optimal_2006}.

%
%
\subsubsection{DoA/ToA EKF}

\MKcolor{Within the DoA and ToA tracking \gls{EKF}, a \gls{CWNA} model is employed for the state-evolution~\cite[Ch.~6.2]{BLK01} in order to track the DoA and ToA estimates. Hence, the state-vector for the $\ell_k$th \gls{AN} can be written as }
\ifonecolumn
\else
    \MKcolor{\begin{align}
    \begin{split}
    \mathbf{s}_{\ell_k}[n] =[ &\tau_{\ell_k}[n],\,\vartheta_{\ell_k}[n],\,\varphi_{\ell_k}[n],\\ &\Delta \tau_{\ell_k}[n],\,\Delta\vartheta_{\ell_k}[n],\,\Delta\varphi_{\ell_k}[n]]\transpose \in \mathbb{R}^{6},
    \end{split}
    \end{align}}
\fi
\MKcolor{where $\varphi_{\ell_k}[n]\in[0,2\pi)$ and $\vartheta_{\ell_k}[n]\in[0,\pi]$ denote the azimuth and co-elevation \gls{DoA} angles at the time-instant $n$, respectively. Similarly, $\tau_{\ell_k}[n]$ denotes the \gls{ToA} at the $\ell_k$th \gls{AN}. Finally, the parameters $\Delta \tau_{\ell_k}[n]$, $\Delta\vartheta_{\ell_k}[n]$, and $\Delta\varphi_{\ell_k}[n]$ denote the rate-of-change of the \gls{ToA} as well as of the arrival-angles, respectively. In addition, let us consider the measurement model presented in \eqref{eq:EKF_fitted_model} and the following linear state evolution model that stems from the assumed \gls{CWNA} model}
\MKcolor{
\begin{align}
	\mathbf{s}_{\ell_k}[n] &= \mathbf{F}\mathbf{s}_{\ell_k}[n-1]+ \mathbf{u}[n], && \mathbf{u}[n] \sim \mathcal{N}(0,\mathbf{Q}[n]),
	\label{eq:state_transition5}
\end{align}}
\MKcolor{where the state transition matrix $\mathbf{F} \in \mathbb{R}^{6 \times 6}$ as well as the covariance matrix of the state-noise $\mathbf{Q}[n] \in \mathbb{R}^{6 \times 6}$ are given by}
\ifonecolumn
\else
\MKcolor{\begin{align}
	    \mathbf{F} &= \begin{bmatrix} \mathbf{I}_{3\times 3} & \Delta t\cdot \mathbf{I}_{3\times 3} \\ \mathbf{0}_{3\times3} & \mathbf{I}_{3\times 3} \end{bmatrix}, \\
	    \mathbf{Q}[n] &= \begin{bmatrix}
        \frac{\sigma_w^2 \Delta t^3}{3} \cdot \mathbf{I}_{3 \times 3}  & \frac{\sigma_w^2 \Delta t^2}{2} \cdot \mathbf{I}_{3 \times 3} \\ \frac{\sigma_w^2 \Delta t^2}{2} \cdot \mathbf{I}_{3 \times 3} & \sigma_w^2 \Delta t \cdot \mathbf{I}_{3 \times 3} \end{bmatrix}.
\end{align}}
\fi

\MKcolor{Here, $\Delta t$ denotes the time-interval between two consecutive estimates. We note that $\mathbf{F}$ and $\mathbf{Q}[n]$ may be found by employing the so-called numerical discretization of the following continuous-time state model~\cite[Ch.2]{HSS11}
\begin{equation}
    \frac{d \mathbf{s}_{\ell_k}(t)}{dt} = \begin{bmatrix} \mathbf{0}_{3 \times 3} & \mathbf{I}_{3 \times 3}\\\mathbf{0}_{3 \times 3} & \mathbf{0}_{3 \times 3}\end{bmatrix}\mathbf{s}_{\ell_k}(t) + \begin{bmatrix}\mathbf{0}_{3 \times 3} \\ \mathbf{I}_{3 \times 3} \end{bmatrix} \mathbf{w}(t),
\end{equation}
where $\mathbf{w}(t) \in \mathbb{R}^{3 \times 1}$ denotes a white-noise process with the diagonal power spectral density $\mathbf{Q}_c = \sigma_w^2\mathbf{I}_{3\times 3}$.
}


The prediction and update equations of the information-form of the \gls{EKF} for the $\ell_k$th \gls{AN} can now be expressed as 
\ifonecolumn
\else
\begin{align} \label{eq:EKF_DoAToA}
    \est{\mathbf{s}}^-_{\ell_k}[n] &= \mathbf{F} \est{\mathbf{s}}^+_{\ell_k}[n-1]\\
	\mathbf{P}^-_{\ell_k}[n] &= \mathbf{F}\mathbf{P}^+_{\ell_k}[n-1]\mathbf{F} \transpose + \mathbf{Q}[n]\label{eq:pred_cov_2}\\
	\mathbf{P}^+_{\ell_k}[n] &= \left(\left(\mathbf{P}^-_{\ell_k}[n]\right)^{-1} + \mathbf{J}_{\ell_k}[n] \right)^{-1}\\
	\est{\mathbf{s}}^+_{\ell_k}[n] &= \est{\mathbf{s}}^-_{\ell_k}[n] + \mathbf{P}^+_{\ell_k}[n] \mathbf{v}_{\ell_k}[n],
\end{align}
\fi
where $\mathbf{J}_{\ell_k}[n] \in \mathbb{R}^{6 \times 6}$ and $\mathbf{v}_{\ell_k}[n] \in \mathbb{R}^{6 \times 1}$ denote the observed \gls{FIM} and score-function of the state evaluated at $\hat{\mathbf{s}}^{-}_{\ell_k}[n]$, respectively. They are found by employing the measurement model 
for the estimated \gls{UL} channel in~\eqref{eq:EKF_fitted_model}, and concentrating the corresponding log-likelihood function wrt. the path weights. In particular, the observed \gls{FIM} and score-function are given by \cite{Ric05,VOK91,CRK2012}
\ifonecolumn
\else
    \begin{align} \label{eq:FIM_scorefunc}
        &\mathbf{J}_{\ell_k}[n] = \frac{2}{\sigma_n^2}\Re\left\{\left(\frac{\partial \mathbf{r}}{\partial \hat{\mathbf{s}}_{\ell_k}^{-\transpose}[n]}\right)\conT \frac{\partial \mathbf{r}}{\partial \hat{\mathbf{s}}_{\ell_k}^{-\transpose}[n]} \right\},\\
        &\mathbf{v}_{\ell_k}[n] = -\frac{2}{\sigma_n^2}\Re\left\{\left(\frac{\partial \mathbf{r}}{\partial \hat{\mathbf{s}}_{\ell_k}^{-\transpose}[n]}\right)\conT \mathbf{r} \right\}.
    \end{align}
\fi
Here, $\mathbf{r} = \mathbf{\Pi}^\bot(\hat{\mathbf{s}}_{\ell_k}^{-}[n]) \mathbf{g}_{\ell_k}[n]$ and $\mathbf{\Pi}^\bot(\hat{\mathbf{s}}_{\ell_k}^{-}[n]) = \mathbf{I} - \mathbf{\Pi}(\hat{\mathbf{s}}_{\ell_k}^{-}[n])$ denotes an orthogonal projection matrix onto the nullspace of $\mathbf{B}(\vartheta,\varphi,\tau)$; see Section \ref{sec:ChannelModel_tracking}. In particular, $\mathbf{\Pi}(\hat{\mathbf{s}}_{\ell_k}^{-}[n]) = \mathbf{B}(\vartheta,\varphi,\tau) \mathbf{B}^\dag(\vartheta,\varphi,\tau)$, where the superscript $\{\cdot\}^\dag$ denotes the Moore-Penrose pseudo-inverse.

\subsubsection{EKF Initialization}
Initial estimates of the \gls{DoA}, \gls{ToA}, and respective rate-of-change parameters are needed for initializing the \gls{EKF} proposed in the previous section. Here, we describe a simple yet reliable approach for finding such initial estimates. In particular, the initial estimates $\hat{\vartheta}_{\ell_k}[0]$, $\hat{\varphi}_{\ell_k}[0]$, and $\hat{\tau}_{\ell_k}[0]$ are found as follows:


\begin{itemize}
    \item Reshape the \gls{UL} channel vector into a matrix\MKcolor{\footnote{\MKcolor{Reshape operator denoted as mat\{$\mathbf{X}, d_1,d_2,\dots,d_q$\} reshapes a given matrix or vector $\mathbf{X}$ into a $d_1 \times d_2 \times \dots \times d_q$ matrix.}}}:
    \begin{equation}
        \bm{\mathcal{H}}_{\ell_k} = \mathrm{mat}\{\mathbf{g}_{\ell_{k}},\mathcal{M}_f,\mathcal{M}_{\mathrm{AN}}\}
    \end{equation}
    \item Multiply $\bm{\mathcal{H}}_{\ell_k}$ with the \gls{EADF} for horizontal and vertical components, and reshape into a 3D matrix:
    \ifonecolumn
\else
        \begin{align}
            \mathbf{A}_H  &= \mathrm{mat}\{\bm{\mathcal{H}}_{\ell_k}\mathbf{G}_{{\ell_k}_H}\conT,\mathcal{M}_f,\mathcal{M}_a,\mathcal{M}_e\}, \\
            \mathbf{A}_V  &= \mathrm{mat}\{\bm{\mathcal{H}}_{\ell_k}\mathbf{G}_{{\ell_k}_V}\conT,\mathcal{M}_f,\mathcal{M}_a,\mathcal{M}_e\}
    \end{align}
    \fi
    
    \item Employ the 3D \gls{FFT} and determine
    \ifonecolumn
\else
        \begin{align}
            \mathbf{B}_H &= |\mathrm{FFT}_{3D}\{\mathbf{A}_H\}|^2,\\
            \mathbf{B}_V &= |\mathrm{FFT}_{3D}\{\mathbf{A}_V\}|^2
        \end{align}
    \fi
    
    \item Find the indices of the largest element of the 3D matrix $\mathbf{B}_H + \mathbf{B}_V$. These indices correspond to the estimates $\hat{\vartheta}_{\ell_k}[0]$, $\hat{\varphi}_{\ell_k}[0]$, and $\hat{\tau}_{\ell_k}[0]$.
\end{itemize}
We note that the initialization method described above is a computationally efficient implementation of the space-time conventional beamformer (deterministic \gls{MLE} for a single path), and it stems from the work in \cite{Ric05,CRK2012}. The initialization of the covariance matrix may be achieved by evaluating the observed \gls{FIM} at $\hat{\mathbf{s}}^+_{\ell_k}[0]$, and using $\mathbf{P}^+_{\ell_k}[0] = (\mathbf{J}_{\ell_k}[0])^{-1}$.

The rate-of-change parameters may be initialized once two consecutive estimates of $[\hat{\vartheta}_{\ell_k}, \hat{\varphi}_{\ell_k}, \hat{\tau}_{\ell_k}]$ are obtained. For example, in order to initialize $\Delta \tau_{\ell_k}$ at $n=2$ the following can be used
\begin{align}
    \Delta \tau_{\ell_k}[2] &= \frac{\tau_{\ell_k}[2]-\tau_{\ell_k}[1]}{\Delta t},\\
    (\mathbf{P}_{\ell_k}^+[2])_{4,4} &= \frac{1}{(\Delta t)^2}\left((\mathbf{P}_{\ell_k}^+[1])_{1,1} + (\mathbf{P}_{\ell_k}^+[2])_{1,1}\right),
\end{align}
where the notation $(\mathbf{A})_{i,j}$ denotes the entry of matrix $\mathbf{A}$ located at the $i$th row and $j$th column.
%
%
\subsection{Positioning and Synchronization EKF at Central Processing Unit}

Next, an algorithm for the simultaneous \gls{UN} positioning and clock synchronization is presented, following the preliminary work by the authors in \cite{werner_joint_2015}. Since in practice every \gls{UN} has an offset in its internal clock wrt. the \glspl{AN}' clocks, it is crucial to track the clock offset of the \gls{UN} in order to achieve reliable \gls{ToA} estimates for positioning. Furthermore, different clock offsets among the \glspl{AN} should be taken into account as well, but that topic is covered in more detail in Section \ref{sec:syncpos}. In this section, we first present the novel EKF solution, called joint DoA/ToA Pos\&Clock EKF, for simultaneous \gls{UN} positioning and clock synchronization for the case when \glspl{AN} are synchronized. Then, a practical and improved initialization method for the presented Pos\&Clock EKF is also proposed. For notational simplicity, we assume below that only a single UN is tracked. However, assuming orthogonal UL pilots, the ToAs and DoAs of multiple UNs can, in general, be estimated and tracked, thus facilitating also simultaneous positioning and clock offset estimation and tracking of multiple devices.

%
\subsubsection{Joint DoA/ToA Pos\&Clock EKF}\label{ssec:joint_doa_toa_ekf}

Within the joint DoA/ToA Pos\&Clock EKF, the obtained ToA and DoA estimates from different LoS ANs are used to estimate the UN position and velocity as well as the clock offset and clock skew of the UN. Thus, the state of the process is defined as
\begin{equation}
\mathbf{s}[n] = [x[n],y[n],v_x[n],v_y[n],\rho[n],\alpha[n]]\transpose \in \mathbb{R}^{6},
\end{equation}
where $\mathbf{p}[n] = [x[n], y[n]]\transpose $ and $ \mathbf{v}[n]=[v_x[n], v_y[n]]\transpose$ are two-dimensional position and velocity vectors of the UN, respectively. Furthermore, the clock offset $\rho[n]$ and the clock skew $\alpha[n]$ of the UN are assumed to evolve according to the clock models in \eqref{eq:clock_offset} and \eqref{eq:clock_skew}.

Let us next assume that the velocity of the UN is almost constant between two consecutive time-steps, being only perturbed by small random changes, i.e., the state evolution model is a \gls{CWNA} model~\cite[Ch.~6.2]{BLK01}. Then, stemming from this assumption and since the clock models for the evolution of the clock offset and skew are linear, a joint linear model for the state transition can be expressed as
\begin{equation}
    \mathbf{s}[n] = \mathbf{F}_{\textrm{UN}}\mathbf{s}[n-1] + \mathbf{w}[n],
    \label{eq:state_transition_eq1}
\end{equation}
where the state transition matrix $\mathbf{F}_{\textrm{UN}} \in \mathbb{R}^{6 \times 6}$ is
\begin{gather}
	\mathbf{F}_{\textrm{UN}} = \begin{bmatrix}
	\mathbf{I}_{2\times2} & \Delta t\cdot\mathbf{I}_{2\times2} & \mathbf{0}_{2\times2}\\
	\mathbf{0}_{2\times2} & \mathbf{I}_{2\times2} & \mathbf{0}_{2\times2}\\
	\mathbf{0}_{2\times2} & \mathbf{0}_{2\times2} & \mathbf{F}_\mathrm{c}
\end{bmatrix},\quad 
\mathbf{F}_\mathrm{c} = 
	\begin{bmatrix}
		1 & \Delta t\\
		0 & \beta
	\end{bmatrix}.\label{eq:state_transition_matrix}
\end{gather}
Here, the process noise is assumed to be zero-mean Gaussian such that $\mathbf{w}[n] \sim \mathcal{N}(0, \mathbf{Q}')$ where the discretized block diagonal covariance $\mathbf{Q}'\in \mathbb{R}^{6 \times 6}$ is given by
\ifonecolumn
\else
    \begin{equation}
        \mathbf{Q}' =  \begin{bmatrix}
            \frac{\sigma_v^2 \Delta t^3}{3} \cdot \mathbf{I}_{2 \times 2}  & \frac{\sigma_v^2 \Delta t^2}{2} \cdot \mathbf{I}_{2 \times 2} & \mathbf{0}_{2 \times 1} & \mathbf{0}_{2 \times 1} \\
            \frac{\sigma_v^2 \Delta t^2}{2} \cdot \mathbf{I}_{2 \times 2} & \sigma_v^2 \Delta t \cdot \mathbf{I}_{2 \times 2} & \mathbf{0}_{2 \times 1} & \mathbf{0}_{2 \times 1} \\
            \mathbf{0}_{1 \times 2} & \mathbf{0}_{1 \times 2} & \frac{\sigma_{\eta}^2 \Delta t^3}{3} & \frac{\sigma_{\eta}^2 \Delta t^2}{2}\\
            \mathbf{0}_{1 \times 2} & \mathbf{0}_{1 \times 2} & \frac{\sigma_{\eta}^2 \Delta t^2}{2} & \sigma_{\eta}^2 \Delta t
        \end{bmatrix}.
        \label{eq:process_noise2}
    \end{equation}
\fi
Here, $\sigma_v$ and $ \sigma_\eta$ denote the \gls{STD} of the velocity and clock skew noises, respectively.

The upper left corner of the state transition matrix $\mathbf{F}_{\textrm{UN}}$ represents the constant movement model of the UN, whereas the matrix $\mathbf{F}_\mathrm{c}$ describes the clock evolution according to the clock models \eqref{eq:clock_offset} and \eqref{eq:clock_skew}. Both presented clock models have been shown to be suitable for clock tracking in \cite{kim_tracking_2012} using practical measurements. Unfortunately, the authors in \cite{kim_tracking_2012} do not provide details on the values for the parameter $\beta$ as determined in their experiments. Although the clock skew is not necessarily completely stationary, the change in the clock skew is relatively slow compared to the clock offset. Therefore, the authors in \cite{kim_tracking_2012} argue that the clock skew can be assumed to be quasi-stationary for long time periods. According to the calculations and observations in \cite{werner_joint_2015}, we employ the value $\beta = 1$ throughout the paper. Thus, the clock skew can eventually be considered as a random-walk process as well.

In contrast to the linear state transition model, the measurement model for the joint DoA/ToA Pos\&ClockEKF is non-linear. For each time step $n$, let us denote the number of \glspl{AN} with a \gls{LoS} condition to the UN as $K[n]$ and the indices of those \glspl{AN} as $\ell_1, \ell_2, \dots, \ell_{K[n]}$. For each LoS-AN $\ell_k$, the measurement equation consists of the azimuth DoA estimate $\hat{\varphi}_{\ell_k}[n] = \varphi_{\ell_k}[n] + \delta \varphi_{\ell_k}[n]$ and the ToA estimate $\hat{\tau}_{\ell_k}[n] = \tau_{\ell_k}[n] + \delta \tau_{\ell_k}[n]$, where $\delta \varphi_{\ell_k}[n]$ and $\delta \tau_{\ell_k}[n]$ denote estimation errors for the obtained azimuth DoA and ToA measurements, respectively. Note that the focus is on 2D position estimation, and thus the estimated elevation angles are not employed by this EKF. The measurements for each AN can thus be combined into a joint measurement equation expressed as 
\begin{equation}
    \mathbf{y}_{\ell_k}[n] = [\hat{\varphi}_{\ell_k}[n], \hat{\tau}_{\ell_k}[n]]\transpose = \mathbf{h}_{\ell_k}(\mathbf{s}[n]) + \mathbf{u}_{\ell_k}[n],
\end{equation}
where $\mathbf{u}_{\ell_k}[n] = [\delta \varphi_{\ell_k}[n], \delta \tau_{\ell_k}[n]]\transpose$ is the zero-mean observation noise with a covariance $\mathbf{R}_{\ell_k}[n] = \mathbb{E}[\mathbf{u}_{\ell_k}[n]\mathbf{u}_{\ell_k}\transpose [n]]$. Furthermore, $\mathbf{h}_{\ell_k}(\mathbf{s}[n]) = [h_{{\ell_k},1}(\mathbf{s}[n]), h_{{\ell_k},2}(\mathbf{s}[n])] \transpose $ is the real-valued and non-linear measurement function that relates the measurement vector $\mathbf{y}_{\ell_k}[n]$ to the UN state through
\ifonecolumn
\else
    \begin{align}
        &h_{{\ell_k},1}(\mathbf{s}[n]) = \textrm{arctan}\left( \frac{\Delta y_{\ell_k}[n]}{\Delta x_{\ell_k}[n]}\right)\\
        &h_{{\ell_k},2}(\mathbf{s}[n]) = \frac{d_{\ell_k}[n]}{c} + \rho [n],\label{eq:toa}
    \end{align}
\fi    
where $\Delta x_{\ell_k}[n] = x[n] - x_{\ell_k}$ and $\Delta y_{\ell_k}[n] = y[n] - y_{\ell_k}$ are distances between the AN ${\ell_k}$ and the UN in $x$- and $y$-direction, respectively. In \eqref{eq:toa}, the two-dimensional distance between the UN and the AN is denoted as $d_{\ell_k} = \sqrt{\Delta x_{\ell_k}^2[n] + \Delta y_{\ell_k}^2[n]}$ and the speed of light is denoted as $c$. Finally, the complete measurement equation containing measurements $\mathbf{y}_{\ell_k}$ from all \gls{LoS}-\glspl{AN} at time step $n$ can be written as
\begin{equation}
    \mathbf{y}[n] = \mathbf{h}(\mathbf{s}[n]) + \mathbf{u}[n],
    \label{eq:measurement_eq1}
\end{equation}
where $\mathbf{y} = [\mathbf{y}_{\ell_1} \transpose,\mathbf{y}_{\ell_2} \transpose,\dots,\mathbf{y}_{\ell_{K[n]}} \transpose] \transpose$ is the collection of measurements and $\mathbf{h} = [\mathbf{h}_{\ell_1} \transpose,\mathbf{h}_{\ell_2} \transpose,\dots,\mathbf{h}_{\ell_{K[n]}} \transpose] \transpose$ is the respective combination of the model functions. Furthermore, the noise $\mathbf{u}[n] \sim \mathcal{N}(0,\mathbf{R})$ with a block diagonal covariance matrix $\mathbf{R} = \textrm{blkdiag}\left(\left[\mathbf{R}_{\ell_1}, \mathbf{R}_{\ell_2}, \dots, \mathbf{R}_{\ell_{K[n]}}\right]\right)$ describes the zero-mean measurement errors for all $K[n]$ \gls{LoS}-\glspl{AN}.

Let us next assume that the initial state $\est{\mathbf{s}}^+[0]$ as well as the initial covariance matrix $\est{\mathbf{P}}^+[0]$ are known. \MKcolor{Furthermore, assuming the linear state transition model and the non-linear measurement model as derived in \eqref{eq:state_transition_eq1} and \eqref{eq:measurement_eq1}, respectively, the well-known Kalman-gain form of the EKF can be applied for estimating the state of the system~\cite{simon_optimal_2006}.} Because of the linear state transition model, the prediction phase of the EKF can be applied in a straightforward manner within the joint Pos\&Clock EKF. \MKcolor{However, the Jacobian matrix $\mathbf{H}$ used in the update phase needs to be evaluated at $\est{\mathbf{s}}^-[n]$ before applying the subsequent equations of the EKF.} It is straightforward to show that after simple differentiation the elements of the Jacobian matrix $\mathbf{H}$ become
\ifonecolumn
\else
    \begin{align}
        \mathbf{H}_{2k-1,1}[n] &= \left[\mathbf{h}_{{\ell_k},1}\right]_{x}(\hat{\mathbf{s}}^-[n]) = -\frac{\Delta \hat{y}_{\ell_k}[n]}{\hat{d}_{\ell_k}^2[n]}\label{eq:jacobian1}\\
        \mathbf{H}_{2k-1,2}[n] &= \left[\mathbf{h}_{{\ell_k},1}\right]_{y}(\hat{\mathbf{s}}^-[n]) =\frac{\Delta \hat{x}_{\ell_k}[n]}{\hat{d}_{\ell_k}^2[n]}\\
        \mathbf{H}_{2k,1}[n] &= \left[\mathbf{h}_{{\ell_k},2}\right]_{x}(\hat{\mathbf{s}}^-[n]) =\frac{\Delta \hat{x}_{\ell_k}[n]}{c\,\hat{d}_{\ell_k}[n]}\\
        \mathbf{H}_{2k,2}[n] &= \left[\mathbf{h}_{{\ell_k},2}\right]_{y}(\hat{\mathbf{s}}^-[n]) =\frac{\Delta \hat{y}_{\ell_k}[n]}{c\,\hat{d}_{\ell_k}[n]}\label{eq:jacobian4}\\
        \mathbf{H}_{2k,5}[n] &= \left[\mathbf{h}_{{\ell_k},2}\right]_{\rho}(\hat{\mathbf{s}}^-[n]) = 1, \label{eq:jacobian5}
\end{align}
\fi
for $k = 1, 2, \dots, {K[n]}$ and zero otherwise~\cite{werner_joint_2015}. In \eqref{eq:jacobian1}-\eqref{eq:jacobian4}, we denote distances between the AN and the predicted UN position in $x$- and $y$-directions as $\Delta \hat{x}_{\ell_k}[n]$ and $\Delta \hat{y}_{\ell_k}[n]$, respectively. Similarly, the notation $\hat{d}_{\ell_k}[n]$ denotes the two-dimensional distance between the $\ell_k$th AN and the predicted UN position. 

At every time step $n$, the two-dimensional UN position estimate is hence obtained as $\est{\mathbf{p}}[n] = [(\est{\mathbf{s}}^+[n])_1, (\est{\mathbf{s}}^+[n])_2]\transpose $ with an estimated covariance found as the upper-left-most $2 \times 2$ submatrix of $\est{\mathbf{P}}^+[n]$. In addition to the UN position estimate, an estimate of the UN clock offset is given through the state variable $(\est{\mathbf{s}}^+[n])_5$ as a valuable by-product.

%
%

\subsubsection{EKF Initialization}\label{ssec:ekf_init}

Initialization of the EKF, i.e., the choice of the mean $\est{\mathbf{s}}^+[0]$ and the covariance $\est{\mathbf{P}}^+[0]$ of the initial Gaussian distribution plays an important role in the performance of the EKF. In the worst case scenario, poorly chosen initial values for the state and covariance might lead to undesired divergence in the EKF whereas good initial estimates ensure fast convergence. 
Here, we propose a practical two-phase initialization method for the Pos\&Clock EKF in which no external information is used besides that obtained through the normal communication process between the UN and \glspl{AN}. The proposed initialization method is illustrated in Fig.~\ref{fig:init_phase}.

In the first phase of initialisation, we determine coarse initial position and velocity estimates of the UN together with their respective covariances which are used, thereafter, as an input to the next phase of the proposed initialization method. In the literature, there are many different methods that can be used to determine such initial position estimates. For example, the authors in \cite{zampella_indoor_2015} used \gls{RSS} measurements to obtain the position estimates whereas in \cite{navarro_toa_2007} the authors used DoA and \gls{ToA} based methods for the UN positioning. The UN could even communicate position estimates that are obtained by the UN itself using, e.g., \gls{GNSS}, but has the disadvantage of increasing the amount of additional communication between the UN and \glspl{AN}, and such an external positioning service is not necessarily always available. In our initialization method, we apply the \gls{CL} method~\cite{bulusu_gps-less_2000} building on the known positions of the \gls{LoS}-\glspl{AN} in order to obtain a rough position estimate for the UN. Thus, the initial position estimate, denoted as $\est{\vec{p}}^+[0] = [\est{x}^+[0], \est{y}^+[0]]\transpose $, can be expressed as
\begin{equation}
  \est{\vec{p}}^+[0] = \frac{1}{K[0]}\sum_{k = 1}^{K[0]} \vec{p}_{\ell_k},
  \label{eq:cl_method}
\end{equation}
where $K[0]$ is the total number of \gls{LoS}-\glspl{AN} and $\vec{p}_{\ell_k}$ denotes the known position of the \gls{LoS}-AN with an index $\ell_k$. Intuitively, \eqref{eq:cl_method} can be understood as the mean of the \gls{LoS}-\glspl{AN}' positions, and depending on the location of the UN relative to the \gls{LoS}-ANs the initial position estimate may be poor. Such coarse initial position estimate can be improved by using a \gls{WCL} method where the weights can be obtained from, e.g., \gls{RSS} measurements \cite{pivato_accuracy_2011}. 

\ifonecolumn
\else
{\linespread{\speciallinespacing}
\begin{figure}
    \centering
    \includegraphics[width=3.12in]{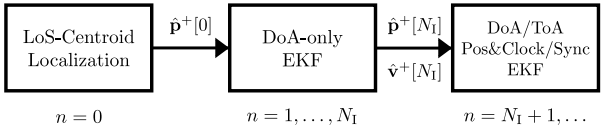}
    \caption{Initialization of position estimate $\est{\vec{p}}^+[n]$ where also a velocity estimate $\est{\vec{v}}^+[n]$ is improved as a by-product in the second phase of the initialization}
    \label{fig:init_phase}
    \vspace{\spaceunderfig}
\end{figure}
}
\fi

Unless the positioning method provides an initial estimate also for the velocity, the EKF can be initialized even with a very coarse estimate. If external information about the environment or device itself is available, a reliable estimate for the velocity can be easily obtained considering, e.g., speed limits of the area where the obtained initial position estimate is acquired. However, since external information is not used in our initialization method, the initial velocity estimate of the UN is set to zero without loss of generality. By combining the initial position and velocity estimates we can determine a reduced initial state estimate $\est{\mathbf{s}}^+[0]$ that can be used as an input for the next phase of the proposed initialization method.

It is also important that the employed initialization method provides not only the state estimate but also an estimate of the covariance. In our method, the uncertainty of the initial position is set to a large value, since the initial position estimate obtained using the CL method might easily be coarse and, therefore, cause divergence in the EKF if a small uncertainty is used. Since the initial velocity is defined without any further assumptions, it is consequential to set the corresponding covariance also to a large value. Hence, by setting the initial covariance to be reasonably large we do not rely excessively on the uncertain initial state. 

However, the initial position estimate obtained using the above initialization procedure may not be accurate enough to ensure reliable convergence in the presented DoA/ToA positioning and synchronization EKF, especially in the sense of using susceptible ToA measurements in the update phase of the filtering.  Therefore, we chose to execute DoA-only EKF, i.e., an EKF where only the azimuth DoA measurements are momentarily used to update the state estimate of the UN \cite{werner_estimating_2013}, in the second phase of the overall initialization procedure. The DoA-only EKF, in which the obtained initial state and covariance estimate are used as prior information, is carried out only for pre-defined $N_\textrm{I}$ iterations. In addition to more accurate position estimate, we can also estimate the UN velocity $\hat{\mathbf{v}}[n] = \left[ \hat{v}_x[n],\hat{v}_y[n]\right]\transpose$ as a by-product in the DoA-only EKF. 

The state estimate obtained from the DoA-only EKF after the $N_\textrm{I}$ iterations can then be used to initialize the joint DoA/ToA Pos\&Clock EKF after the state has been extended with the initial \gls{UN} clock parameters. In the beginning, the clock offset can be limited to a fairly low value by simply communicating the time from one of the \gls{LoS}-\glspl{AN}. Thereafter, the communicated time can be used to set up the clock within the UN. Typically, manufacturers report the clock skew of their oscillators in parts per million (ppm). Based on the results achieved in the literature, e.g., in \cite{kim_tracking_2012,cristea_fingerprinting_2013,kohno_remote_2005} the clock skew of the UN can be initialized to $\est{\alpha}^+[0] = 25$~ppm with a \gls{STD} of a few tens of ppm~\cite{werner_joint_2015}. Finally, the extended state and covariance that contain also the necessary parts for the clock parameters can be used as prior information for the actual DoA/ToA Pos\&Clock EKF as well as for the yet more elaborate DoA/ToA Pos\&Sync EKF proposed next in Section~\ref{sec:syncpos} for the case of unsynchronized network elements.

\section{Cascaded EKFs for Joint UN Positioning and Network Clock Synchronization} \label{sec:syncpos}

In the previous section, we assumed that the clock of a UN is unsynchronized with respect to \glspl{AN} whereas the \glspl{AN} within a network are mutually synchronized. In this section, we relax such an assumption, by considering unsynchronized rather than synchronized \glspl{AN}.
For mathematical tractability and presentation simplicity, we assume that the \glspl{AN}' clocks within a network are, however, phase-locked, i.e., the clock offsets of the ANs are static. It is important to note that this assumption does not imply the same clock offsets between the \glspl{AN}, leaving thus a clear need for network synchronization. In the following, an EKF for both joint UN positioning and network synchronization is proposed. The issue of propagating a universal time within a network is also discussed.

\subsection{Positioning and Network Synchronization EKF at Central Unit}

In general, the proposed EKF for simultaneous UN positioning and network synchronization, denoted as a joint DoA/ToA Pos\&Sync EKF, is an extension to the previous joint DoA/ToA Pos\&Clock EKF where also the mutual clock offsets of the \gls{LoS}-\glspl{AN} are tracked using the available ToA measurements. An augmented state where also the clock offsets of all \gls{LoS}-\glspl{AN} at time step $n$ are considered can now be expressed as 
\begin{equation}
\begin{aligned}
\vec{s}[n] = [\vec{s}_{\textrm{UN}}\transpose [n], \rho_{\ell_1}[n], \cdots, \rho_{\ell_{K[n]}}[n]]\transpose \in \mathbb{R}^{6+K[n]},
\end{aligned}
\label{eq:final_state}
\end{equation}
where $\vec{s}_{\textrm{UN}}\transpose [n] = [x[n],y[n],v_x[n],v_y[n],\rho[n],\alpha[n]]$ is the same state vector containing the position and velocity of the UN as well as the clock parameters of the UN clock as presented in Section~\ref{ssec:joint_doa_toa_ekf}. Furthermore, the clock offset of the \gls{LoS}-AN with an index $\ell_k$ where $k \in 1,2, \ldots, K[n]$ is denoted in the augmented state as $\rho_{\ell_k}[n]$. Here, all the clock offsets are interpreted relative to a chosen reference AN~clock.

Since the clocks within \glspl{AN} are assumed to be phase-locked, we can now write the clock offset evolution model for the AN with an index $\ell_k$ as
\begin{equation}
    \rho_{\ell_k}[n] = \rho_{\ell_k}[n-1] + \delta_{\rho}[n],
    \label{eq:an_clocks}
\end{equation}
where $\delta_{\rho} \sim \mathcal{N}(0,\sigma_{\rho}^2)$ denotes the zero-mean Gaussian noise for the clock offset evolution. Using the model~\eqref{eq:an_clocks} for the clock offsets and assuming the same motion model~\eqref{eq:state_transition_eq1}, we can write then a linear transition model for the state~\eqref{eq:final_state} within DoA/ToA Pos\&Sync EKF such that 
\begin{align}
    \mathbf{s}[n] = \mathbf{F}[n]\mathbf{s}[n-1] + \mathbf{w}[n],
    \label{eq:transition_model1}
\end{align}
where $\mathbf{w}[n] \sim \mathcal{N}(0,\mathbf{Q}[n])$ denotes the zero-mean distributed noise with the following covariance 
\begin{equation}
    \mathbf{Q}[n] = \left[ \begin{array}{cc}
       \mathbf{Q}' & \mathbf{0}_{K[n] \times K[n]} \\
        \mathbf{0}_{K[n] \times K[n]} & \sigma_{\rho}^2 \Delta t \cdot \mathbf{I}_{K[n] \times K[n]}\\
    \end{array}\right],
\end{equation}
where $\mathbf{Q}'$ is the same covariance as in~\eqref{eq:process_noise2}. Furthermore, the augmented state transition matrix $\mathbf{F}[n] \in \mathbb{R}^{(6+K[n]) \times (6+K[n])}$ can be written as

\begin{equation}
    \mathbf{F}[n] = \begin{bmatrix} \mathbf{F}_{\textrm{UN}} & \mathbf{0}_{6\times K[n]} \\
    \mathbf{0}_{K[n]\times 6} & \mathbf{I}_{K[n] \times K[n]}\\
    \end{bmatrix},
    \label{eq:state_transition_matrix2}
\end{equation}
where the matrix $\mathbf{F}_{\textrm{UN}} \in \mathbb{R}^{6 \times 6}$ represents the same state transition matrix for the UN state as in~\eqref{eq:state_transition_matrix}. The identity matrix in the lower-right corner of the state transition matrix $\mathbf{F}[n]$ represents the assumed clock offset evolution for the \glspl{AN} as presented in~\eqref{eq:an_clocks}.

Next, due to the mutually unsynchronized ANs, the measurement equation related to ToA in~\eqref{eq:toa} needs to be revised accordingly. Thus, by adding the clock offset of the considered \gls{LoS}-\gls{AN} to the earlier ToA measurement equation in \eqref{eq:toa}, we can write new measurement equations as
\ifonecolumn
\else
    \begin{align}
        &h_{{\ell_k},1}(\mathbf{s}[n]) = \textrm{arctan}\left( \frac{\Delta y_{\ell_k}[n]}{\Delta x_k[n]}\right)\\
        &h_{{\ell_k},2}(\mathbf{s}[n]) = \frac{d_{\ell_k}[n]}{c} + \rho[n] + \rho_{{\ell_k}}[n],\label{eq:toa2}
    \end{align}
\fi
where $\rho_{\ell_k}[n]$ denotes the clock offset of the \gls{LoS}-\gls{AN} with an index ${\ell_k}$. Furthermore, the measurement equations in \eqref{eq:toa2} for each LoS-AN can be combined into the similar measurement model as in~\eqref{eq:measurement_eq1} such that
\begin{equation}
    \mathbf{y}[n] = \mathbf{h}(\mathbf{s}[n]) + \mathbf{u}[n],
    \label{eq:meas12}
\end{equation}
where $\mathbf{u}[n] \sim \mathcal{N}(0, \mathbf{R})$ is the measurement noise with covariance $\mathbf{R} = \textrm{blkdiag}\left(\left[\mathbf{R}_{\ell_1}, \mathbf{R}_{\ell_2}, \dots, \mathbf{R}_{\ell_{K[n]}}\right]\right)$ obtained from the DoA/ToA tracking phase.

\MKcolor{In the following, we apply the Kalman-gain form of the EKF to the presented models \eqref{eq:transition_model1} and \eqref{eq:meas12} in order
to obtain the joint DoA/ToA Pos\&Sync EKF.} Since our measurement model contains now also the clock offset parameters for the LoS-\glspl{AN}, we need to modify the Jacobian matrix $\mathbf{H}$ in \eqref{eq:jacobian1}-\eqref{eq:jacobian5} by adding the corresponding elements for each LoS-\glspl{AN}, namely
\begin{equation}
    \mathbf{H}_{2k,6+k}[n] = [\mathbf{h}_{{\ell_k},2}]_{\rho_{\ell_k}}(\hat{\mathbf{s}}^-[n]) = 1,
\end{equation}
where $k \in 1, 2, \dots, K[n]$, and zeros elsewhere to complete the matrix. 

In the beginning of the filtering, e.g., when a UN establishes a connection to the network, we use the same initialization method as proposed earlier in Section~\ref{ssec:ekf_init}. Thereafter, the \gls{UN} position and clock offset estimates at time step $n$ are obtained as $[(\hat{\mathbf{s}}^+[n])_1,(\hat{\mathbf{s}}^+[n])_2] \transpose$ and $(\hat{\mathbf{s}}^+[n])_5$, respectively, with estimated covariances found as respective elements of the matrix $\hat{\mathbf{P}}^+$. Since the proposed DoA/ToA Pos\&Sync EKF also tracks now the clock offsets of the LoS-\glspl{AN}, the estimated clock offsets for each LoS-AN are given through the state estimates $(\hat{\mathbf{s}}^+[n])_{6+k}$ where $k \in 1,2,\dots, K[n]$. These obtained UN and LoS-\glspl{AN} clock offset estimates can be used thereafter in network synchronization.

In order to be able to track the offsets of LoS-\glspl{AN} properly and define synchronization within an unsynchronized network, we need to choose one of the LoS-ANs as a reference AN. Since the ToA measurements are not used in the earlier proposed initialization phase, the reference AN can be chosen to be, e.g., the closest AN to the UN when the initialization phase is completed and the ToA measurements are started to be used in positioning and clock offset tracking. This implies that in the initial phase, $\rho_{\ell_1}[n] = 0$, assuming that the AN $\ell_1$ refers to the reference AN. Thus, synchronization can be achieved within the network wrt. the reference AN by communicating the clock offset estimates for the LoS-\glspl{AN} and~\mbox{the UN}.

%
%
\subsection{Propagation of Universal Network Time} \label{sec:network_time}

In the case of tracking only one UN at the time, synchronization of the network is done with respect to the reference time obtained from a chosen reference AN. However, when we apply the proposed method for multiple UNs simultaneously we have to consider how to treat clock offset estimates that have different time references \cite{KM08,RV15}. In general, it is realistic to assume that there are a large number of UNs connected to a network, and thus tracked, simultaneously. Therefore, we can obtain clock offset estimates for numerous \glspl{AN} within a network using the proposed method such that the clock offsets for each AN have been estimated using different reference times. If this information is stored and available in a central unit, the relative offsets of these \glspl{AN} can be estimated easily and the network can be thereafter synchronized wrt. any of these \glspl{AN}. 

However, storing the clock offset information increases the computational load and the use of memory capacity in the central node of a network and, therefore, alternative approaches how to utilize the estimated clock offsets can be considered. As an alternative approach, the same relative offset information could be used also as a prior information for the clock offset estimation in the case of new UNs. If this information is available in the beginning of tracking a new UN, it would most probably speed up convergence in the proposed EKF and even improve the clock offset estimate of the UN. Further aspects related to establishing and propagating a universal network time is an interesting and important topic for our future research.


\section{Numerical Evaluations and Analysis}\label{sec:results}

In this section, comprehensive numerical evaluations are carried out to illustrate and quantify the achievable device positioning performance using the proposed methods. The evaluations are carried out in an urban outdoor environment, adopting the METIS Madrid grid model~\cite{metis_simulation_2013}, while the deployed UDN is assumed to be operating at the 3.5~GHz band. We specifically focus on the connected car use case \cite{osseiran_scenarios_2014,ngmn_alliance_5g_2014,5g-ppp_2015} where cars are driving through a city with velocities in the order of \SI{50}{\kilo\meter\per\hour}. In all our evaluations, we deploy comprehensive map and ray tracing based propagation modeling~\cite{metis_channel_2015} such that the modeling of the incoming UL reference signals in different ANs is as realistic as possible and explicitly connected to the environment and map. Further details of the environment and evaluation methodologies are given below. In general, the performances of the proposed joint DoA/ToA Pos\&Clock and Pos\&Sync EKFs, as well as the DoA-only based EKF implemented for reference, are evaluated and reported.


\subsection{Simulation and Evaluation Environment}

In the following, a detailed description of the employed simulation environment is presented. First, the structure and properties of the Madrid grid environmental model~\cite{metis_simulation_2013} that stems from the METIS guidelines are described with an essential accuracy. After that, details of the used ray tracing channel model are discussed and finally a realistic motion model for the UN is presented. 

\subsubsection{Madrid map}

Outdoor environment has a huge impact not only on constraining the UN movement but also on wireless communications, especially, when modelling the radio signal propagation within a network.  The Madrid map, which refers to the METIS Madrid grid environmental model, is considered as a compromise between the existing models like Manhattan grid and the need of characterising dense urban environments in a more realistic manner~\cite{metis_simulation_2013}. For evaluating and visualizing the positioning performance, we used a two-dimensional layout of the Madrid map as illustrated in Fig.~\ref{fig:madrid_map}.

\iftwocolumn
{\linespread{\speciallinespacing}
\begin{figure}
    \centering
    \includegraphics[scale=0.5]{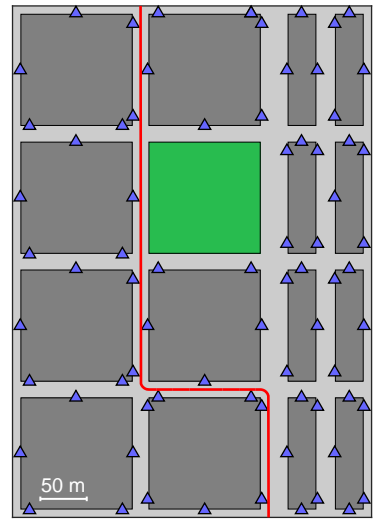}
    \caption{Madrid map with example \acrfull{AN} deployment (blue triangles) and \acrfull{UN} trajectory (red path).}
    \label{fig:madrid_map}
    \vspace{\spaceunderfig}
\end{figure}
}
\fi

In the connected car application, we model only the necessary parts of the Madrid map based on the METIS guidelines~\cite{metis_simulation_2013}, i.e., the indoor model as well as minor details like bus stops and metro entrances are ignored during the process. The majority of the Madrid map is covered with square and rectangle shaped building blocks as represented in Fig.~\ref{fig:madrid_map} with dark gray color. Square blocks have both dimensions equal to \SI{120}{m} whereas length and width of the other building blocks are \SI{120}{m} and \SI{30}{m}, respectively. The height of the buildings range from \SI{28}{m} to \SI{52.5}{m}. In addition to the buildings, the map contains also a park with the same dimensions as square shaped buildings, and it is located almost in the middle of the map. The rest of the map is determined to be roads and sidewalks, but for the sake of simplicity, sidewalks are not illustrated in Fig.~\ref{fig:madrid_map}. In general, these \SI{3}{m} wide sidewalks are surrounding every building in the map, but in our visualizations they are represented as a part of the roads. Road lanes are \SI{3}{m} wide and they are accompanied by \SI{3}{m} wide parking lanes, except the vertical lanes in the widest Gran Via road on the right side of the park. Thus, the normal roads are \SI{18}{m} wide in our evaluations and visualization containing also the sidewalks. Special Grand Via road consists of three lanes in both directions, where the lanes in different directions are separated by \SI{6}{m} wide sidewalk. The parallel road on the right side of Gran Via road is called Calle Preciados and it is defined as a \SI{21}{m} sidewalk in the METIS guidelines~\cite{metis_simulation_2013}. Despite the fact that the sidewalks are illustrated as a part of the roads in Fig.~\ref{fig:madrid_map}, we do not allow vehicles to move on these sidewalks.

\subsubsection{Channel and antenna models} \label{sssec:channel_models}

We employ the ray tracing as well as the geometry-based stochastic channel models described in \cite{metis_channel_2015,3GPP_channel}. In particular, the ray tracing channel model is employed in order to model the propagation of the \gls{UL} reference signals that are exploited by the proposed \glspl{EKF} for \gls{UN} positioning as realistically as possible. \MKcolor{The employed ray tracing implementation} takes into account the 3D model of the Madrid grid when calculating the reflected and diffracted paths between the UN and ANs. The diffracted paths are given according to the Berg's model \cite{metis_channel_2015}. Moreover, the antennas composing the arrays at the ANs are assumed to observe the same directional channel, and thus a single-reference point at the AN's location is used in calculating the ray tracing channels. The effect of random scatterers is also modeled according to the METIS guidelines \cite{metis_channel_2015} with a density of~\SI{0.01}{scatterers\per\meter\squared}.

The geometry-based stochastic channel model (GSCM) \cite{metis_channel_2015,3GPP_channel} is used in this paper in order to model uncoordinated interference. In particular, the interferers are randomly placed on a disk-shaped area ranging between \SI{200}{m} and \SI{500}{m} away from the ANs receiving the UL reference signals. A density of \SI{1000}{interferers\per\kilo\meter\squared} is used and their placement follows a Poisson point process. The channels among the interferers' and multiantenna ANs are calculated according to the GSCM, and used to calculate a spatially correlated covariance matrix at the receiving ANs. This is done for all subcarriers modulated by the UL reference signals. Such a covariance matrix is then used to correlate a zero-mean complex-circular white-Gaussian distributed vector for each UL transmission. This approach of modeling uncoordinated interference is similar to that in \cite{3GPP_mimo_simulations}.

The multiantenna transceivers at the ANs are assumed to have a cylindrical geometry; see Fig. \ref{fig:cylindrical_array}. In particular, the cylindrical arrays are comprised of $10$ dual-polarized patch-elements, and thus $20$ output ports, while the height of the AN antenna system is assumed to be \SI{7}{m}. The beampatterns of the patch-elements are taken from \cite{3GPP_channel}. The patch-elements are placed along two circles, each with an inter-element distance of $\lambda/2$. The vertical separation between the two circles is also $\lambda/2$. Moreover, the circles have a relative rotation/shift of $2\pi/10$. Note that the \gls{EADF} given in Section \ref{sec:system_model} is found and calculated for this antenna array. Finally, the UN employs a vertically-oriented dipole, at height \SI{1.5}{m}, while the interferers are equipped with randomly-oriented dipoles.

\iftwocolumn
{\linespread{\speciallinespacing}
\begin{figure}
    \centering
    \includegraphics[scale=0.5]{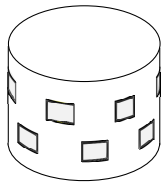}
    \caption{Illustration of the 3D array geometry employed at the ANs. Cylindrical arrays comprising $10$ dual-polarized 3GPP patch-elements are used. The array elements are placed along two circles each of which comprising $5$ patch-elements.}
    \label{fig:cylindrical_array}
    \vspace{\spaceunderfig}
\end{figure}
}
\fi 

\subsubsection{\gls{UN} motion model}
In order to demonstrate that the proposed system is capable of positioning \glspl{UN} with realistic time-varying velocities as well as time-varying accelerations, we assume that the \glspl{UN} are moving in vehicles on trajectories such as the one depicted in \rfig{fig:madrid_map}. \MKcolor{On the straight parts of the trajectory}, the vehicle is assumed to accelerate up to a maximum velocity of $v_{\text{m}} = \SI{50}{km/h}$, whereas all turns are performed with a constant velocity of $v_{\text{t}} = \SI{20}{km/h}$. The time-varying acceleration from $v_{\text{t}}$ to $v_{\text{m}}$ and the time-varying deceleration from $v_{\text{m}}$ to $v_{\text{t}}$ are modelled according to polynomial models stemming from real-life traffic data described in~\cite{akcelik_acceleration_1987}. The polynomial model in~\cite{akcelik_acceleration_1987} makes it possible to create acceleration profiles with varying characteristics. In this work, we generate profiles that follow from the estimation of acceleration time and distance as described in~\cite{akcelik_acceleration_1987}. The resulting acceleration profile for one \gls{UN} route depicted in \rfig{fig:madrid_map} and velocities $v_{\text{m}} = \SI{50}{km/h}$ and $v_{\text{t}} = \SI{20}{km/h}$ is shown in \rfig{fig:acceleration_profile}.

\subsubsection{5G Radio Interface Numerology and System Aspects}

The 5G UDN is assumed to adopt OFDMA based radio access with \SI{75}{kHz} subcarrier spacing, \SI{100}{MHz} carrier bandwidth and 1280 active subcarriers. This is practically 5 times up-clocked radio interface numerology, compared to 3GPP LTE/LTE-Advanced radio network, and is very similar to those described, e.g.,\cite{Lahetkangas13,mogensen_5g_2013}. The corresponding radio frame structure incorporates subframes of length \SI{0.2}{ms}, which include 14 OFDM symbols. This is also the basic time resolution for UL reference signals. In the upcoming evaluations, both continuous and sparse UL reference signal subcarrier allocations are deployed, for comparison purposes, while the UN transmit power is always \SI{0}{dBm}. In both reference signal cases, 256 pilot subcarriers are allocated to a given UN which are either continuous (\SI{19.2}{MHz}) or sparse over the whole carrier passband width of \SI{96}{MHz}. Building on the \gls{UL} reference signals, \gls{LS}-based multicarrier-multiantenna channel estimator is adopted in all \glspl{AN}. Also, two different ISDs of \SI{50}{m} and \SI{25}{m} in the UDN design are experimented.

In the evaluations, we assume that the UL reference signals of all the UNs within a given AN coordination area are orthogonal, through proper time and frequency multiplexing. However, also co-channel interference from uncoordinated UNs is modeled as explained in Section \ref{sssec:channel_models}. Assuming a typical noise figure of \SI{5}{dB}, the signal-to-interference-and-noise ratio (SINR) at the AN receiver ranges between \SI{5}{dB} and \SI{40}{dB}, depending on the locations of the target UN and interfering UNs on the map. 

\iftwocolumn
{\linespread{\speciallinespacing}
\begin{figure}
    \centering
    \includegraphics[width=3.12in]{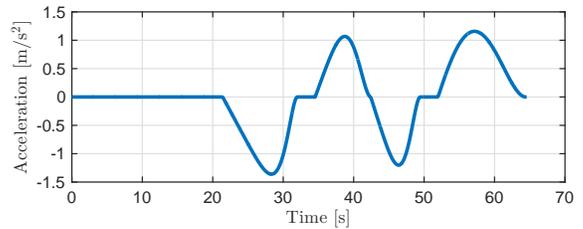}
    \caption{Acceleration profile for the example \gls{UN} trajectory shown in \rfig{fig:madrid_map}.}
    \label{fig:acceleration_profile}
    \vspace{\spaceunderfig}
\end{figure}
}
\fi

In general, all the EKFs are updated only once per \SI{100}{ms}, to facilitate realistic communication of the azimuth DoA and ToA measurements from involved ANs to the central processing unit. \MKcolor{In order to first analyze the effects of the different UL pilot allocations and AN ISDs on DoA and ToA estimation EKF as well as on positioning EKFs, only $K[n]=2$ closest LoS-ANs are fused. After that, we also evaluate the performance of the proposed positioning methods with other realistic numbers of available LoS-ANs while taking into account possible imperfect LoS-detection. This is done for the scenario of sparse pilot allocation and the ISD of \SI{50}{m}.}

\subsection{DoA and ToA Estimation}\label{ssec:doa_toa_estimates}

\iftwocolumn
{\linespread{\speciallinespacing}
\begin{figure}
    \centering
    \includegraphics[height=2.9in, trim=1.2cm 1.2cm 0.7cm 0cm, clip]{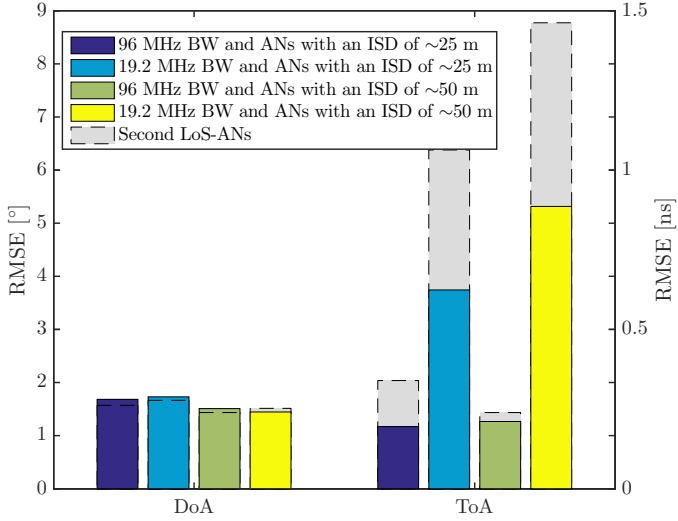}
    \caption{The average RMSEs for the estimated azimuth DoA and ToA at the closest LoS-ANs (colored bars) and the second LoS-ANs (gray bars) along 15 different random routes through the Madrid map.}
    \label{fig:doa_toa_results}
    \vspace{\spaceunderfig}
\end{figure}
}
\fi

\MKcolor{In order to evaluate first the accuracy of the DoA and ToA tracking in the individual ANs using the proposed DoA/ToA EKF, the RMSEs for both estimates are illustrated in Fig.~\ref{fig:doa_toa_results}, averaged across 15 random routes taken through the Madrid map.} Each colored bar represents a different network configuration used in the evaluations for the LoS-ANs that are the closest to the UN whereas bars with a gray colour represent the respective results for the second closest LoS-ANs.

As expected, the ToA estimation and tracking is more accurate when the UL beacons are transmitted using the wider \SI{96}{MHz} bandwidth and a sparse subset of subcarriers than using the narrower \SI{19.2}{MHz} bandwidth due to enhanced time-domain resolution. Decreasing the ISD leads to better ToA estimates due to higher average SINRs at the ANs, especially when using the narrow bandwidth while the difference is not so significant in the case of the \SI{96}{MHz} bandwidth. 

The accuracy of the azimuth DoA estimates, in turn, is generally very high. In general, since the variance of the azimuth angle estimation is always smaller, the more coplanar geometry between the TX and RX we have, the average accuracy of the DoA estimates does not substantially vary between the different ISDs, or between the closest and second closest ANs. This is indeed because the geometry of more far away UNs is more favorable for azimuth angle estimation. In general, one can conclude that excellent ToA and DoA estimation and tracking accuracy can be obtained using the proposed EKF.


\subsection{Positioning, Clock and Network Synchronization}

\MKcolor{Next, the performance of the proposed DoA/ToA Pos\&Clock and Pos\&Sync EKFs is evaluated by tracking UNs moving through the earlier described Madrid map, again with 15 randomly drawn trajectories. }
Each generated route starts from an endpoint of a road on the map with some pre-determined initial velocity. Thereafter, the motion of the UN is defined according to the presented motion model. The routes are defined to end when the UN crosses 6 intersections on the map. For the sake of simplicity, the UN is moving in the middle of the lane. In all the evaluations, the update period of the positioning and synchronization related EKFs at the central processing unit is only every $500$th radio sub-frame, i.e., only every \SI{100}{ms}. This reflects a realistic situation such that the DoA and ToA measurements of individual ANs can be realistically communicated to and fused at the central unit.

\iftwocolumn
{\linespread{\speciallinespacing}
\begin{figure}
    \centering
    \includegraphics[height=2.9in, trim=1.0cm 1.2cm 0.7cm 0cm, clip]{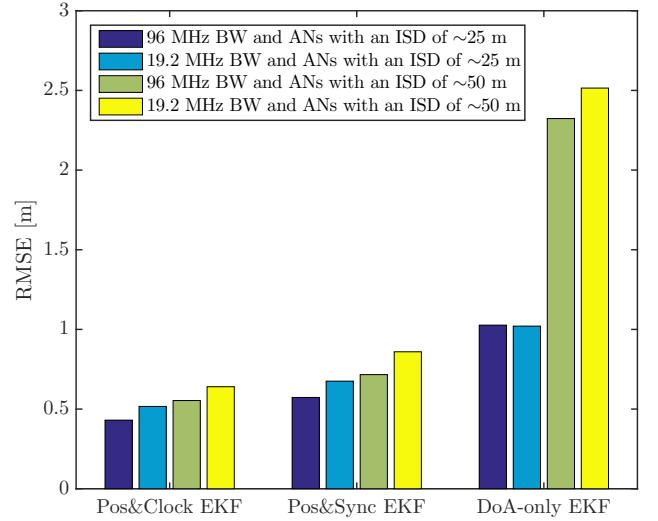}
    \caption{Positioning RMSEs for all tracking methods and with different simulation numerologies, \MKcolor{along 15 different random routes taken through the Madrid map.}}
    \label{fig:positioning_results}
    \vspace{\spaceunderfig}
\end{figure}
}
\fi

Before the actual evaluations, in case of unsynchronized ANs, we initialize the clock offsets of all unsynchronized \glspl{AN} within a network according to $\rho_{\ell_k}[0] \sim \mathcal{N}(0,\sigma_{\rho,0}^2)$ with $\MKcolor{\sigma_{\rho,0} = \SI{100}{\micro\second}}$ as motivated in Section~\ref{ssec:ekf_init}. 
Whenever a new UN is placed on the map, we initialize the UN position estimate $\hat{\mathbf{p}}[0]$ using the CL method within the proposed initialization process. In our evaluations, covariance of the initialized position estimate is defined as a diagonal matrix $\sigma_{p,0}^2 \cdot \mathbf{I}_{2 \times 2}$ where $\sigma_{p,0}$ is set to a large value using the distance between the initial position estimate and current LoS-\glspl{AN}. Furthermore, we set the initial velocity according to $\left[\hat{v}_x[0],\hat{v}_y[0]\right]\transpose \sim \mathcal{N}(0,\sigma_{v,0}^2 \cdot \mathbf{I}_{2 \times 2})$ with quite large \gls{STD} of $\sigma_{v,0} = \SI{5}{m/s}$ based on the earlier discussion in Section~\ref{ssec:ekf_init}. The initial estimates that we have determined for the UN so far are then used as a prior for the DoA-only EKF within the proposed initialization method. The DoA-only EKF is executed for $N_{\textrm{I}}=20$ iterations to initialize the more elaborate EKFs, in terms of position and velocity. Thereafter, we need to initialize also the necessary clock parameters in order to use the actual DoA/ToA Pos\&Clock and Pos\&Sync EKFs. As motivated in Section~\ref{ssec:ekf_init}, we set the clock offset and skew for the UNs according to $\rho[0] \sim \mathcal{N}(0,\sigma_{\rho,0}^2)$ where $\sigma_{\rho,0} = \SI{100}{\micro\second}$, and $\alpha[0] \sim \mathcal{N}(\mu_{\alpha,0},\sigma_{\alpha,0}^2)$ where $\mu_{\alpha,0} = \SI{25}{ppm}$ and $\sigma_{\alpha,0} = \SI{30}{ppm}$, respectively. In addition to setting the initial clock parameters, we also choose the reference AN to be the closest LoS-AN to the UN before we start to use the final DoA/ToA Pos\&Clock and Pos\&Sync EKFs for the positioning and network synchronization purposes. The same values are also used for the initialization of the DoA-only EKF that is used as a comparison method for the proposed more elaborate EKFs.
\iftwocolumn
{\linespread{\speciallinespacing}
\begin{figure}
    \centering
    \includegraphics[height=2.9in, trim=1.0cm 1.2cm 0.7cm 0cm, clip]{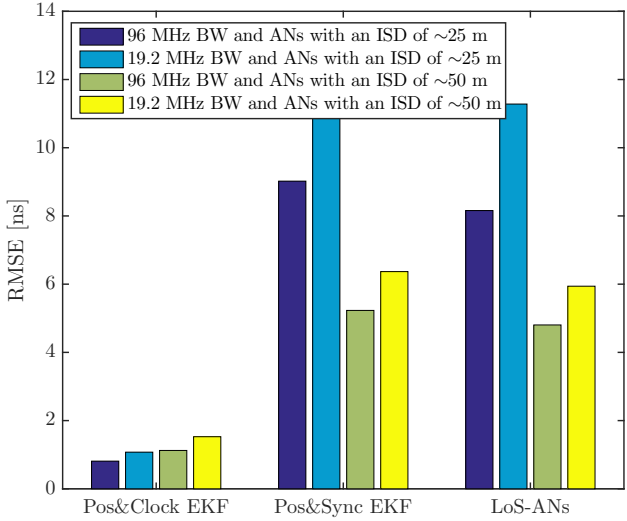}
    \caption{The average RMSEs for the UN clock offset estimates along 15 different random routes through the Madrid map, with synchronous (left) and unsynchronous (middle) ANs. Also shown are the respective RMSEs for the LoS-ANs mutual clock offset estimates (right).}
    \label{fig:clock_results}
    \vspace{\spaceunderfig}
\end{figure}
}
\fi
Furthermore, we set the \gls{STD} of the clock skew driving noise in the clock model \eqref{eq:clock_skew} to $\sigma_{\eta} = 6.3\cdot10^{-8}$ based on the measurement results in \cite{kim_tracking_2012}. However, the \gls{STD} of the clock skew within the EKF is increased to $\sigma_{\eta} = 10^{-4}$ since it leads to a much better overall performance especially when the clock offset and clock skew estimates are very inaccurate, e.g., in the initial offset tracking phase. Since we assume that the UN is moving in a vehicle in an urban environment, we set the \gls{STD} of UN velocity to $\sigma_v = \SI{3.5}{m/s}$. 

Position and clock offset tracking performance of the proposed cascaded \gls{DoA}/\gls{ToA} Pos\&Clock and Pos\&Sync EKFs in comparison to the \gls{DoA}-only \gls{EKF} is illustrated in  Figures~\ref{fig:positioning_results}-\ref{fig:clock_results}, where each color represents a different simulation setup used in the evaluations. In contrast to the classical \gls{DoA}-only \gls{EKF}, the \glspl{RMSE} obtained using the \gls{DoA}/\gls{ToA} Pos\&Clock and Pos\&Sync EKFs are partitioned according to network synchronization assumptions. Furthermore, we also analyse the accuracy of the \gls{UN} clock offset estimates in both synchronized and phase-locked networks. For the sake of simplicity, we fuse the azimuth \gls{DoA} and \gls{ToA} estimates at each EKF update period of \SI{100}{ms} only from two closest \gls{LoS}-\glspl{AN}. The first $10$ EKF iterations (one second in real time) after the initialization procedure are excluded in the RMSE calculations, to avoid any dominating impact of the initial estimates on the tracking results.

Based on the obtained positioning results that are illustrated in Fig.~\ref{fig:positioning_results} the proposed Pos\&Clock and Pos\&Sync EKFs significantly outperform the earlier proposed DoA-only EKF in all considered evaluation scenarios. In particular, an impressive sub-meter positioning accuracy, set as one core requirement for future 5G networks in \cite{5g-ppp_2015}, is achieved by the both proposed methods in all test scenarios, and they even attain positioning accuracy below \SI{0.5}{m} in RMSE sense with the \SI{96}{MHz} bandwidth and ISD of around \SI{25}{m}. An unfavourable and known feature of the DoA-only EKF is that its performance degrades when the geometry of the two LoS-ANs and the UN resembles a line. Since the proposed Pos\&Clock and Pos\&Sync EKFs use also the ToA estimates for ranging, they do not suffer from such disadvantageous geometries. 

In the case of a synchronized network, the Pos\&Clock EKF achieves highly accurate synchronization between the unsynchronized UN and network with an RMSE below \SI{2}{\nano\second} in every test scenario as illustrated in Fig.~\ref{fig:clock_results}. Since the presented ToA estimation errors in Fig.~\ref{fig:doa_toa_results} are between \SI{0.1}{\nano\second} and \SI{1.5}{\nano\second}, these propagate very well to the achievable clock offset tracking in the fusion EKF. Interestingly the high initial clock offset STD of \SI{100}{\micro s} is, in general, improved by 5 orders of magnitude.

Investigating next the achievable clock-offset estimation accuracy with unsynchronized ANs in Fig.~\ref{fig:clock_results} (Pos\&Sync EKF), we can clearly observe that overall the performance is somewhat worse than in the corresponding synchronous case. Furthermore, network densification from ISD of \SI{50}{m} down to \SI{25}{m} actually degrades the UN clock offset estimation accuracy to some extent. These observations can be explained with the assumed motion model and how the clock offsets of the LoS-ANs are initialized within the EKF. When the UN is moving at the velocity of \SI{50}{km/h}, each LoS-AN along the route, with ISD of \SI{25}{m}, is in LoS condition with the UN only \SI{1.8}{\second} and, therefore, we can obtain only 18 DoA/ToA measurements in total from each LoS-AN due to assumed update period of \SI{100}{\milli\second}. Therefore, the Pos\&Sync EKF can be executed a lower number of iterations for a given LoS-AN pair, compared to the network with \SI{50}{m} ISD. This, in turn, means that the initial more coarse clock offset estimates of the individual LoS-ANs have relatively higher weight, through the measurement equation~\eqref{eq:toa2}, to the UN clock offset estimate in the network with ISD of \SI{25}{m}. However, even in the presence of unsynchronized network elements, UN clock offset can be estimated with an accuracy of around \SIrange{5}{10}{ns}, as depicted in Fig.~\ref{fig:clock_results}. Furthermore, the results in Fig.~\ref{fig:clock_results} (LoS-ANs) also demonstrate that highly accurate estimates of the mutual clock offsets of the ANs can be obtained using the proposed cascaded Pos\&Sync EKF. \MKcolor{In particular, the proposed method provides clock offset estimates of the network elements which are significantly more accurate than the expected \SI{0.5}{\micro s} timing misalignment requirement for future 5G small-cell networks~\cite{mogensen_5g_2013}.}

\MKcolor{In addition, the performance of the proposed positioning and synchronization methods was further evaluated with other realistic values of available LoS-ANs as well as under imperfect LoS-detection using the \SI{96}{MHz} bandwidth scenario with the \gls{ISD} of around \SI{50}{m}. First, the positioning and synchronization accuracy is evaluated using the azimuth DoA and ToA measurements either from the closest LoS-AN only or from the three closest LoS-ANs, i.e., $K[n]=1$ and $K[n]=3$, respectively. Second, the imperfect LoS-detection scheme is considered where azimuth DoA and ToA measurements from three closest ANs are fused such that one of the three ANs is NLoS-AN with a probability of 10\%, thus increasing the level of realism in the performance evaluations. The obtained positioning and UN clock offset estimation results from the comprehensive numerical evaluations are depicted in Fig.~\ref{fig:additional_pos_results} and Fig.~\ref{fig:additional_clk_results}, respectively.}

\iftwocolumn
{\linespread{\speciallinespacing}
\begin{figure}[t]
    \centering
    \includegraphics[height=2.9in, trim=1.0cm 1.2cm 0.7cm 0cm, clip]{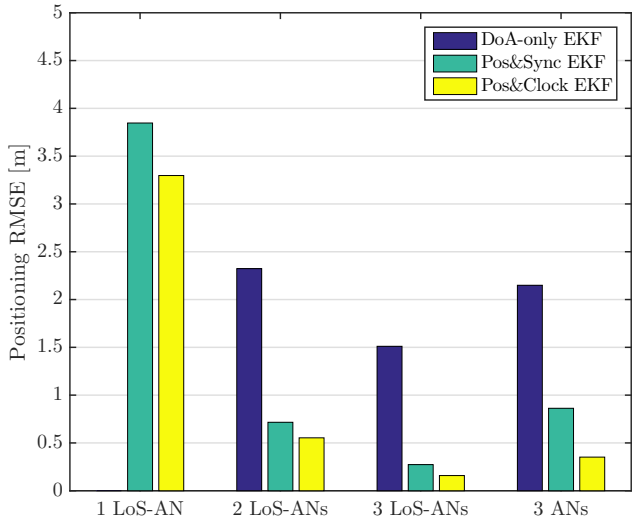}
    \caption{\MKcolor{Positioning RMSEs for all tracking methods with different number of \gls{LoS}-\glspl{AN} and under imperfect \gls{LoS}-detection (denoted as 3 \glspl{AN}), along 15 different random routes taken through the Madrid map.}}
    \label{fig:additional_pos_results}
     \vspace{\spaceunderfig}
\end{figure}
}
\fi

\MKcolor{Based on the obtained positioning results in Fig.~\ref{fig:additional_pos_results}, the positioning performance improves when the azimuth \gls{DoA} and \gls{ToA} measurements are fused from three closest \gls{LoS}-\glspl{AN} compared to the earlier scenario, where the measurement from two closest \gls{LoS}-\glspl{AN} were fused. In particular, positioning accuracy of less than \SI{30}{cm} can be achieved with the proposed methods even under unsynchronized network elements when $K[n]=3$. Such a positioning accuracy is considered as a minimum requirement for, e.g., future autonomous vehicles and \gls{ITS}~\cite{5g-ppp_automotive_2015}. Interestingly, in the case of $K[n]=1$, the performance of the proposed methods is still relatively good although naturally somewhat lower compared to $K[n]=2$ and $K[n]=3$ cases, while more classical DoA-only \gls{EKF} needs the azimuth \glspl{DoA} at least from two \glspl{AN}. Moreover, despite the small and expected degradation of performance due to fusing incorrect azimuth DoA and ToA estimates from NLoS-ANs, in the case of incorrect LoS-detection, the proposed methods are still able to provide sub-meter positioning accuracy also in a such realistic scenario as illustrated in the rightmost bar set of Fig.~\ref{fig:additional_pos_results}. }

\MKcolor{In addition, the obtained UN clock offset estimation results in Fig.~\ref{fig:additional_pos_results} demonstrate that the clock offset estimation performance also improves when the three closest \gls{LoS}-\glspl{AN} are available compared to the scenario, where the measurement from the two closest \gls{LoS}-\glspl{AN} were fused. In the case of $K[n]=1$, the Pos\&Clock EKF outperforms the Pos\&Sync EKF as expected, since imperfect convergence of the UN clock offset estimate in the beginning of a trajectory accumulates throughout the trajectory in the unsynchronized network. In general, rapid and unfavourable handovers which may occur, e.g., in intersections, degrade the performance of clock offset estimation of both UN and ANs within the Pos\&Sync EKF, especially when $K[n]=1$. Despite the imperfect LoS-detection, the proposed methods are able to provide highly accurate UN clock offset estimates as depicted in the rightmost bar set of Fig.~\ref{fig:additional_pos_results}.}
%

The behaviour and performance of both the joint DoA/ToA Pos\&Sync EKF and the DoA-only EKF in tracking with different simulation configurations are further visualized through the videos that can be found on-line at \texttt{\url{http://www.tut.fi/5G/TWC16/}}.

\iftwocolumn
{\linespread{\speciallinespacing}
\begin{figure}[t]
    \centering
    \includegraphics[height=2.9in, trim=1.0cm 1.2cm 0.7cm 0cm, clip]{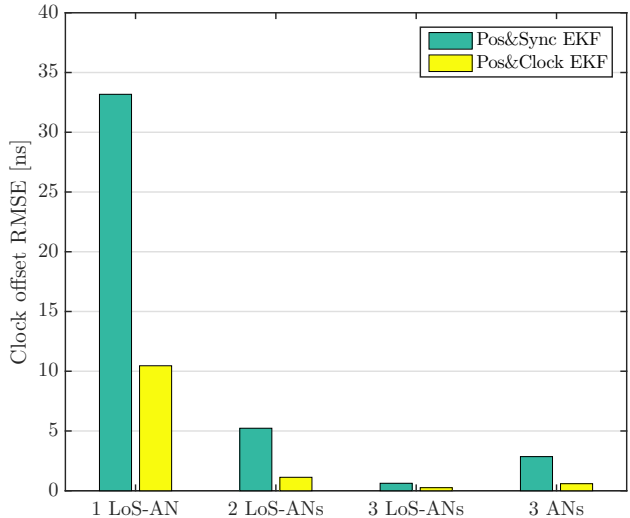}
    \caption{\MKcolor{The average RMSEs for the UN clock offset estimates along 15 different random routes through the Madrid map, with different number of \gls{LoS}-\glspl{AN} and under imperfect \gls{LoS}-detection (denoted as 3 \glspl{AN}).}}
    \label{fig:additional_clk_results}
     \vspace{\spaceunderfig}
\end{figure}
}
\fi

\section{Conclusion}
\label{sec:conclusion}

\MKcolor{In this article, we addressed high-efficiency device positioning and clock synchronization in 5G radio access networks where all the essential processing is carried out on the network side such that the power consumption and computing requirements at the user devices are kept to a minimum.} First, a novel EKF solution was proposed to estimate and track the DoAs and ToAs of different devices in individual ANs, using UL reference signals, and building on the assumption of multicarrier waveforms and antenna arrays. Then, a second novel EKF solution was proposed, to fuse the DoA and ToA estimates from one or more LoS-ANs into a device position estimate, such that also the unavoidable clock offsets between the devices and the network, as well as the mutual clock offsets between the network elements, are all taken into account. Hence, the overall solution is a cascaded EKF structure, which can provide not only highly efficient device positioning but also valuable clock synchronization as a by-product. Then, comprehensive performance evaluations were carried out and reported in 5G \gls{UDN} context, with realistic movement models on the so-called Madrid grid incorporating also full ray tracing based propagation modeling. The obtained results clearly indicate and demonstrate that sub-meter scale positioning and tracking accuracy of moving devices can be achieved using the proposed cascaded EKF solutions even under realistic assumptions. Moreover, network synchronization in the nano-second level can also be achieved by employing the proposed EKF-based scheme. Our future work will focus on extending the proposed solutions to 3D positioning, as well as exploiting the highly accurate positioning information in mobility management and location-based beamforming in 5G networks.

\frenchspacing
\bibliographystyle{IEEEtran}
\bibliography{Sync&Pos_EKF}

\end{document}